\documentclass[%
preprint,
nofootinbib,
longbibliography,
 amsmath,amssymb,
 aps,
]{revtex4-2}

\usepackage{graphicx}
\usepackage{dcolumn}
\usepackage{bm}

\usepackage{subcaption}
\usepackage{xcolor}
\usepackage{ulem}
\usepackage{witharrows}

\usepackage{bbold}
\usepackage{cancel} 
\usepackage{slashed}
\usepackage{natbib}
\usepackage{enumitem} 

\graphicspath{ {./Plots/} }

\newcommand{\st}{{\sin{\theta}}}
\newcommand{\stsq}{{\sin^2{\theta}}}
\newcommand{\sph}{{\sin{\phi}}}
\newcommand{\sphsq}{{\sin^2{\phi}}}
\newcommand{\ct}{{\cos{\theta}}}
\newcommand{\ctsq}{{\cos^2{\theta}}}
\newcommand{\cph}{{\cos{\phi}}}
\newcommand{\cphsq}{{\cos^2{\phi}}}

\begin{document}

\preprint{HIP-2025-11/TH}

\title{Axion Superradiance in Dipole Magnetic Fields of Pulsars}

\author{Topi Sirkiä}
\email{topi.sirkia@helsinki.fi}
\affiliation{Department of Physics, University of Helsinki, 
                      P.O.Box 64, FI-00014 University of Helsinki, Finland}
\affiliation{Helsinki Institute of Physics, 
                      P.O.Box 64, FI-00014 University of Helsinki, Finland}

\author{Matti Heikinheimo} 
\email{matti.heikinheimo@helsinki.fi}
\affiliation{Department of Physics, University of Helsinki, 
                      P.O.Box 64, FI-00014 University of Helsinki, Finland}
\affiliation{Helsinki Institute of Physics, 
                      P.O.Box 64, FI-00014 University of Helsinki, Finland}
                      
\author{Kimmo Tuominen}
\email{kimmo.i.tuominen@helsinki.fi}
\affiliation{Department of Physics, University of Helsinki, 
                      P.O.Box 64, FI-00014 University of Helsinki, Finland}
\affiliation{Helsinki Institute of Physics, 
                      P.O.Box 64, FI-00014 University of Helsinki, Finland}

\date{April 5. 2025}

\begin{abstract}
We consider constraints on the axion-photon coupling by superradiance due to a plasma instability in the magnetospheres of millisecond pulsars. We compute the growth rate of a superradiant axion cloud in a dipole magnetic field, and give a semi-analytical formula for the superradiance rate for the lowest state. By requiring the associated instability time to be longer than the characteristic age of the supermassive black-widow millisecond pulsar PSR J0952–0607, we examine the pulsar-timing array constraints on axions of mass $\sim10^{-12}\text{ eV}$. We find that competitive axion bounds from plasma instabilities are unlikely unless a new high spin pulsar is discovered.
\end{abstract}

\maketitle
\newpage

\section{\label{sec:level1}Introduction}

Ultralight fields such as axions~\cite{Peccei:1977hh,OHare:2024nmr,Marsh:2015xka,Hui:2016ltb,DiLuzio:2020wdo,Sikivie:2020zpn}, dark photons~\cite{Holdom:1985ag,Fabbrichesi:2020wbt} and other axion-like particles (ALPs) are prominent in many well-motivated extensions of the Standard Model (SM). Axions and ALPs appear in the Peccei-Quinn solution of the strong CP problem \cite{Peccei:1977hh,Weinberg:1977ma} and in string theories respectively, while dark photons appear in the simplest extensions of the SM by new gauge interactions. The up-to-date constraints on these particles can be found in~\cite{AxionLimits}. While these fields are constrained by various cosmological observations at wavelengths corresponding to near-galactic distance scales~\cite{Xue:2024zjq,Fedderke:2019ajk,Gan:2023swl} as well as by laboratory searches~\cite{ADMX:2024xbv,Adair:2022rtw,CAST:2017uph}, the wavelengths of astrophysical scales have only recently started to become constrained~\cite{Reynes:2021bpe,Ning:2024eky}. This is largely due to improvement of radio telescopes.

Recently the potential of neutron stars as ultralight axion detectors has been exploited in various ways. For instance, dark matter axion conversion into photons in neutron star (NS) magnetospheres could lead to observable radio signals~\cite{Leroy:2019ghm} which would also exhibit time variations possibly observable by pulsar timing~\cite{Battye:2023oac}. In additional, it has been noted that the strong unscreened electromagnetic fields in pulsar polar caps are expected to source large axion clouds via the coupling $aF\tilde{F}\sim a\vec{E}\cdot\vec{B}$ \cite{Prabhu:2021zve}. These clouds can then emit a broadband or resonant radio flux which could be detected by radio telescopes such as the Square-Kilometer Array (SKA)~\cite{Witte:2021arp,Witte:2024akb,Noordhuis:2023wid}. The constraints obtained are strongest for axions slightly heavier, $10^{-8}  \text{ eV}\lesssim \mu \lesssim 10^{-5} \text{ eV}$, than those which are relevant for stellar superradiance, $10^{-12}  \text{ eV}\lesssim \mu \lesssim 10^{-11} \text{ eV}$. Together the two show strong synergy across a wide range of masses. 

It was shown in 1980 by Detweiler~\cite{PhysRevD.22.2323} that a scalar field in a background Kerr metric of a rotating blackhole is superradiantly unstable. This leads to the growth of black hole (BH) quasi-bound states analogous to those of the hydrogen atom, leading further to what is now called a gravitational atom~\cite{Arvanitaki:2009fg}. Superradiance-induced spindown in BHs has been successfully used to place bounds on ultralight particles by spin measurements. In particular, the lack of black holes in areas of the $(J,M)$-plane could be attributed to superradiance, and the non-observation of such gaps in the plane leads to bounds on ultralight particles whose de Broglie wavelengths match the size of the BH. Such analyses have been done for both stellar remnant BHs~\cite{Witte:2024drg,Baryakhtar:2020gao,Stott:2020gjj} and supermassive BHs~\cite{Hoof:2024quk,Unal:2020jiy}. Additionally, the growth of an axionic cloud leads to many observables such as gravitational wave signals induced by self-interactions \cite{Omiya:2024xlz,Collaviti:2024mvh}, black hole polarimetry~\cite{Gan:2023swl,Chen_2022} and bursts of light~\cite{Boskovic:2018lkj,Ikeda_2019}. However, these bounds are generally not robust as the spindown rates and their derivatives are not known for black holes, but rather only indirectly and model-dependently inferred from accretion disk emission and black hole jets. On the other hand, spin measurements can be made at extremely high precision for neutron stars~\cite{harding2013neutronstarzoo}. 

Superradiance (SR) in general rotating objects has been known for a long time~\cite{zel1971generation}. Zel'dovich showed already in 1972 that a rotating conductive cylinder amplifies electromagnetic waves~\cite{zel1972amplification}. It is now well understood that no horizon akin to black holes is required for SR to occur, but only some dissipative mechanism~\cite{Brito_2020}. Thus SR occurs also in stars, as pointed out in~\cite{Cardoso_2015}. A handful of preliminary bounds have been obtained for ultralight particles from SR in neutron stars, where the dissipative mechanism is supplied by for instance finite-conductivity in the surface~\cite{Cardoso_2017}, the bulk conductivity of the magnetosphere~\cite{Day_2019} or by absorption due to interactions such as the Yukawa coupling~\cite{Kaplan:2019ako}. The spin-down of stars due to SR can then be probed by pulsar-timing arrays. SR has also been considered in other hypothetical compact objects such as boson stars~\cite{Chang:2024xjp,Siemonsen_2021,gao2023bosonstarsuperradiance}. Despite these main results and some development in methods of computing SR rates in stellar media~\cite{Chadha-Day:2022inf}, superradiance in stars is still an emerging tool for constraining ultralight particles.

In this paper, we apply the plasma-induced superradiance scenario proposed in Ref.~\cite{Day_2019} and compute SR for a realistic dipole magnetic field configuration. Using observations of a millisecond pulsar (MSP) with high-spin ($>700$ Hz), we learn that MSPs do not give competitive bounds. This is due to the small size (PSR J1748-2446ad) and anomalously low magnetic field (PSR J0952-0607), as the rate is fairly sensitive both on the mass and radius of the star and its magnetic field strength. Thus, taking into account realistic magnetic field configurations in neutron stars seems to considerably weaken superradiance prospects. However, we postulate that as radio telescopes improve, future observations of high-spin MSPs can make plasma-induced superradiance a viable tool in constraining ultra-light particles. We also note that while we focus on pulsar timing exclusively in this paper, it will be interesting to consider also the light signal from axions that are produced in the cloud, as well as possible polarization signals.

In a plasma background the photon obtains an effective mass $\omega_p$. If this mass is large enough, the superradiant scattering can be effectively blocked~\cite{Spieksma:2025sda}. We discuss how $\omega_p$ is affected by the modeling of the magnetosphere. In this regard our main conclusion is that quantitative features of the effect of $\omega_p$ are very sensitive to the details of the dynamics and the superradiant mechanism we consider in this paper remains a viable possibility.

The paper is organized as follows: In Sec.~\ref{sec:theory} we briefly introduce the calculational framework we use and explicit definitions and formulas needed for the evaluation of the final results. In Sec.~\ref{sec:results} we present our main results and apply them to constrain axion parameter space using observations of MSPs and in Sec.~\ref{sec:plasmaeffects} we discuss the conditions under which photon plasma mass can efficiently suppress superradiant scattering in the magnetosphere. In Sec.~\ref{sec:checkout} we summarize and discuss further prospects of this work. Detailed intermediate results are provided in Appendices.

\section{Theoretical framework and analysis}
\label{sec:theory}

\subsection{Superradiance rate in magnetospheres}
\label{sec:theorySR}

Rotational superradiance generally requires three ingredients which are rotation, bound states and dissipation~\cite{Brito_2020}. In the case of a neutron star the bound states arise by the approximate Schwarzschild geometry of the neutron star exterior.
This gives rise to hydrogen-like bound state configurations of the axion field labeled by mode numbers $n,l,m$. In more detail, the Klein-Gordon equation in this case separates into radial and angular parts with the ansatz $\phi_n\equiv \sum_{lm}\phi_{nlm}=\frac{1}{r}\sum_{lm}Y_{lm}\Phi_{nl}$. The axion configuration with definite  
mode numbers ($n$, $l$, $m$) is then given as $\phi_{nlm}=\frac{1}{r}Y_{lm}\Phi_{nl}$. In the case of long axion wavelengths, or equivalently $\mu r_s \lesssim R \mu \ll 1$, where $r_s$ is the Schwarzschild radius of the star, one finds that the radial function $\Phi_{nl}(r)$ satisfies a Schrödinger-like equation leading to the hydrogen-like solution with
\begin{align}
    \Phi_{nl}=\mu^{1 / 2} \alpha_{nl}^{1 / 2} \sqrt{\frac{n!}{2(n+2 l+1)!(n+l+1)}} e^{-x / 2} x^{l+1} L_n^{2 l+1}(x), \quad \omega_{nl}^2=\mu^2\left(1-\frac{\alpha_{nl}^2}{4}\right),
\end{align}
where
    $x=r \mu \alpha_{nl}$ and $\alpha_{nl}={2\mu M}/(l+n+1)$ is the gravitational coupling and we have set $G\equiv 1$.

This {\em axion cloud} draws energy from the rotational energy of the magnetosphere via dissipative dynamics provided by the finite conductivity of the neutron star magnetosphere: The axion field interacts with photons via the axion-photon coupling and subsequently scatters off the rotating magnetosphere extracting its energy due to the non-hermitean dynamics associated with the dissipative medium. Via another axion-photon conversion this energy is then deposited back to the axion cloud. 

To describe this dynamics, we outline the superradiance rate calculation of Ref. \cite{Day_2019}. The Lagrangian of an axion coupled to a plasma is given by
\begin{equation}
\mathcal{L} \supset \sqrt{-g}\left[\frac{1}{2} \partial_\mu \phi \partial^\mu \phi-\frac{\mu^2}{2} \phi^2-\frac{1}{4} F_{\mu \nu} F^{\mu \nu}-\frac{g_{a \gamma \gamma}}{4} \phi F_{\mu \nu} \tilde{F}^{\mu \nu}-A_\mu j^\mu\right],
\label{eq:axionL}
\end{equation}
where the metric is a flat, $\tilde{F}^{\mu \nu}=\epsilon^{\mu \nu \gamma \delta}F_{\gamma \delta}/2$ is the dual of the electromagnetic field strength tensor, $g_{a \gamma \gamma}$ is the axion-photon coupling and $j^\mu=\sigma F^{\mu \nu} u_\nu+\rho u^\mu$ is the current in the plasma. Quantities $\rho, \sigma$ are the charge density and conductivity of the plasma and $u^\mu$ is the fluid four-velocity which obeys $u^\mu u_\mu=1$. 

The equations of motion arising from the Lagrangian in Eq.~\eqref{eq:axionL} are difficult to solve even numerically and even in the limit $g_{\rm{a}\gamma\gamma}=0$.  The analysis of Ref.~\cite{Day_2019} is done using perturbation theory in both the conductivity $\sigma$ and the axion-photon coupling $g_{\rm{a}\gamma\gamma}$. We note that smaller conductivities typically correspond to longer neutron star lifetimes \cite{Brambilla:2015vta}, which turn out to lead to stronger constraints. Larger conductivities also somewhat counter-intuitively experience a suppression in the superradiance rate, as found in Ref. \cite{Cardoso_2017}. Thus we follow the perturbative approach of Ref.~\cite{Day_2019}

The linearized equations of motion are written in a matrix form as \cite{Day_2019}
\begin{align}
\left[H_F+V_A+V_{a \gamma \gamma}\right]\left(\begin{array}{l}
|\phi\rangle \\
\left|A^0\right\rangle \\
|\mathbf{A}\rangle
\end{array}\right)=\omega^2\left(\begin{array}{l}
|\phi\rangle \\
\left|A^0\right\rangle \\
|\mathbf{A}\rangle
\end{array}\right),
\end{align}
where the free Hamiltonian is $H_F={\rm{diag}}(-d^2/dr_\ast^2+U(r),-\nabla^2,-\nabla^2)$
and the perturbations are given by the matrices
\begin{align}
V_{a \gamma \gamma} & =i g_{a \gamma \gamma}\left(\begin{array}{ccc}
0 & \mathbf{B}({\bf r}) \cdot \hat{p} & -\omega \mathbf{B}({\bf r}) \\
\mathbf{B}({\bf r}) \cdot \hat{p} & 0 & 0 \\
\omega \mathbf{B}({\bf r}) & 0 & 0
\end{array}\right),\,\,
\label{Vay}
\end{align} 
and 
\begin{align}
V_A & =i \sigma(\hat{x})\left(\begin{array}{ccc}
0 & 0 & 0 \\
0 & \mathbf{u}({\bf r}) \cdot \hat{p} & -\omega \mathbf{u}({\bf r}) \\
0 & \hat{p} & -\omega-\mathbf{u}({\bf r}) \times ( \hat{p} \times \cdot )
\end{array}\right) . \label{Va}
\end{align}
Note that $V_{a\gamma\gamma}$ is hermitian while $V_A$ is not. The non-hermiticity is related to dissipation and thus the superradiant scattering effect is completely encoded in Eq.~\eqref{Va}. Using the standard machinery of perturbation theory, one finds the corrected eigenfrequencies $\omega+\delta\omega$ for the states $\left|\phi_{l m n}\right\rangle$ with the result~\cite{Day_2019}
\begin{align}
\delta \omega_{nlm}&=\frac{\pi^2}{8 \omega_{l n}} \sum_{l_1,l_2, m_1,m_2} \sum_{\lambda_1, \lambda_2} \notag \\ & \left\langle\phi_{l m n}\left|V_{a \gamma \gamma}\right| A_{l_1 m_1}^{({\lambda_1)}}\right\rangle\left\langle A_{l_1 m_1}^{({\lambda_1)}}\left|V_A\right| A_{l_2 m_2}^{({\lambda_2)}}\right\rangle\left\langle A_{l_2 m_2}^{({\lambda_2)}}\left|V_{a \gamma \gamma}\right| \phi_{l m n}\right\rangle. \label{deltawnlm}
\end{align}
The photon field has time dependence $\sim\exp(-i\omega t)$ and in spherical coordinates the modes of the photon field are defined by the expansion 
\begin{align}
    A^{(\lambda)}_{\mu,lm}(r,\theta,\phi)&=A_l(\omega r)Y_{lm}(\theta,\phi)\epsilon^{(\lambda)}_\mu,
\end{align}
where we have given the photon states in the complete basis of spherical harmonics $Y_{lm}(\theta,\phi)$ and radial functions $A_l (\omega r)$, which can be solved from the Laplace's equation for the photon field. 
With the assumption of infinitely conducting star, the solution is~\cite{Day_2019} 
\begin{align}
&A_{l}(\omega r)=\frac{1}{N_{l}(\omega R)} \sqrt{\frac{2 \omega}{\pi}}\big(y_{l}(\omega R) j_{l}(\omega r)-j_{l}(\omega R) y_{l}(\omega r)\big), \label{Al}
\end{align}
where $N_l(\omega R)=\left(j_{l}^2(\omega R)+y_{l}^2(\omega R)\right)^{1 / 2} $.
The superradiance rate $\Gamma_{\rm{SR}}$ is then given as the imaginary part of the energy correction. Hence, one needs to compute these matrix elements for the given potentials $V_{a\gamma\gamma}$ and $V_{A}$, i.e. for a given magnetic field $\mathbf{B}({\bf{r}})$ and flow field $\mathbf{u}({\bf{r}})$. 

\subsection{Matrix elements for a dipole magnetic field}

In Ref.~\cite{Day_2019} superradiance was demonstrated in the simple case of a magnetic field in the $\hat{\bf{z}}$-direction. This turns out to approximately capture the essential features of the system in the $l=|m|=1$ case, as the axion cloud is localized close to the equator, where the physical magnetic field is approximately in the $\hat{\bf{z}}$-direction. However, to extract more precise phenomenological predictions one must model the magnetic field more carefully. We therefore assume the more realistic dipole magnetic field
\begin{equation}
\mathbf{B}(r,\theta,\phi)=B_0\big(\frac{R}{r}\big)^3(2 \cos \theta \hat{\boldsymbol{r}}+\sin \theta \hat{\boldsymbol{\theta}}),
\label{eq:DipoleField}
\end{equation}
where $B_0$ is the surface magnetic field. The fluid flow is taken to be azimuthal, rotating with the star with angular velocity $\Omega$. Thus, $\mathbf{u}(r,\theta,\phi)=\Omega r \sin{\theta}\hat{\boldsymbol{\phi}}$.

The above magnetic field is the case of the inclined rotator model where the magnetic and rotational axes of the star align~\cite{Rezzolla_2001}. The more complicated magnetic field structure allows more complicated mode mixing between the different photon polarizations. Additionally, in contrast to the result of Ref.~\cite{Day_2019}, the dipolar field now properly depicts the formation of higher excited cloud states which are no longer concentrated on the equator of the star. The price to pay is that the number of required matrix elements increases and they become more cumbersome to evaluate. The superradiance rate is also expected to be slightly lower due to the $r^{-3}$ suppression~\cite{Day_2019}. In Eq.~\eqref{deltawnlm} the axion-photon conversion matrix elements are Hermitian while the photon-photon elements are dissipative and thus non-Hermitian. 

We need to compute the spatial integrals in the expressions
$\langle\phi_{nlm}|V_{a \gamma \gamma}| A_{l^{\prime} m^{\prime}}^{(\lambda)}\rangle$ 
and $\langle A_{l m}^{(\lambda)}\left|V_A\right| A_{l^{\prime} m^{\prime}}^{(\lambda)}\rangle$
appearing in Eq.~\eqref{deltawnlm}. The summation over Minkowski spacetime components is resolved by using the explicit expressions for the perturbation potentials, Eqs.~\eqref{Vay} and \eqref{Va}, and Cartesian polarisation vectors $\epsilon^{(\lambda)}_\mu=\delta^\lambda_\mu$. Then the computation proceeds using the photon field mode functions $A_{l}(\omega r)Y_{lm}(\theta,\phi)$ as given above and the axion bound state solution $\phi_{nlm}(r,\theta,\phi)$ given in Sec.~\ref{sec:theorySR}.

The radial and angular integrals separate and to treat them, we define
\begin{align}
    I[l_1,l_2,l_3]\equiv & \int_0^\infty dx\ e^{\frac{x}{2}}x^{l_1+1}L_n^{2l_2+1}(x)j_{l_3}(\frac{x}{\alpha_{nl}}), \notag \\
    A[l,m,l',m',f]\equiv &\int d\Omega Y_{lm}^* f(\theta ,\phi)Y_{l'm'} \label{Eq:auxdefs}\\
    S[l_1,l_2,l_3,m_1,m_2,m_3]\equiv & \sqrt{(2l_1+1)(2l_2+1)}\begin{pmatrix}
        l_1&l_2 & l_3 \\
        -m_1&m_2 & m_3
    \end{pmatrix}\begin{pmatrix}
        l_1&l_2 & l_3 \\
        0&0&0
    \end{pmatrix},\notag
\end{align}
where $L^l_n$ is the associated Laguerre polynomial and $j_l$ is the spherical Bessel function. The functions $f(\theta,\phi)$ appearing in the definition of the function $A[l,m,l',m',f]$ are specified in Appendix~\ref{app:functions}. We have used the approximation $A_l(\omega r)\approx \sqrt{\frac{2\omega}{\pi}}j_l(\omega r)$, valid when $\omega R\ll 1$. This is valid for the region where the superradiance condition $\Omega \gtrsim \mu \sim \omega$ is fulfilled as $R^{-1}>\Omega$.

With these ingredients, and assuming $\omega R\ll 1$, we find the four axion-photon matrix elements to be
\begin{align}
    \left\langle\phi_{nlm}\left|V_{a \gamma \gamma}\right| A_{l^{\prime} m^{\prime}}^{(0)}\right\rangle &= 2(-1)^{-m}g_{\rm{a}\gamma\gamma}B_0R^3\mathcal{N}_\phi (\omega \mu\alpha_{nl})^{\frac{1}{2}}\Big(  \omega\mu\alpha_{nl}I[l-2,l,l'-1]\notag \\ &+(\frac{m'}{2}-l'-1)(\mu\alpha_{nl})^2 I[l-3,l,l'])S[l,l',1,m,m',0]\notag \\ 
    &+\frac{c_{l'm'}}{\sqrt{2}}(\mu\alpha_{nl})^2I[l-3,l,l']S[l,l',1,m,m',-1]\Big),
    \notag \\
    \left\langle\phi_{l m n}\left|V_{a \gamma \gamma}\right| A_{l^{\prime} m^{\prime}}^{(i)}\right\rangle&=(-1)^{-m} g_{\rm{a}\gamma\gamma}B_0R^3\mathcal{N}_\phi (\mu\alpha_{nl}\omega)^{\frac{3}{2}}\mathcal{I}^{(i)}_\Omega I[l-2,l,l'],
    \label{phiVayA}
\end{align}
with 
$\mathcal{N}_\phi\equiv \sqrt{n!}[\pi(n+2 l+1)!(n+l+1)]^{-1 / 2}$, $c_{lm}\equiv\sqrt{(l-m)(l+m+1)}$ and
\begin{align}
    \mathcal{I}^{(1)}_\Omega&=- \sqrt{\frac{3}{2}}i\left(S[l,l',2,m,m',-1]-S[l,l',2,m,m',1]\right),\notag \\
    \mathcal{I}^{(2)}_\Omega&=\sqrt{\frac{3}{2}}\left(S[l,l',2,m,m',-1]+S[l,l',2,m,m',1]\right),\notag \\
    \mathcal{I}^{(3)}_\Omega&= -2iS[l,l',2,m,m',0].
\end{align}
The main difference with the example calculation in~\cite{Day_2019} is that now the axion mixes not only with the $A^{(0,3)}_{lm}$ components of the photon but also with the $A^{(1,2)}_{lm}$ components. This means that a priori we have to consider plasma interactions $\left\langle A_{l m}^{(\lambda_1)}\left|V_A\right| A_{l^{\prime} m^{\prime}}^{(\lambda_2)}\right\rangle$ between all photon polarizations. Additionally, the angular dependence of the dipole magnetic field leads to more complicated selection rules. 
We provide explicit expressions for these matrix elements in Appendix~\ref{app:matrixes}.

\section{Results}
\label{sec:results}

\subsection{Leading Order Semi-Analytical Formula}
While the intermediate sum in Eq.~\eqref{deltawnlm} is in principle over all $lm$-states of the intermediate photons, the leading-order contribution to the SR rate can be computed for a given axion state by considering the intermediate state with lowest $l$-value allowed by the selection rules arising from the angular integral or Wigner $3j$-symbol. However, due to the amount of matrix elements and their complexity, the form for the leading order contribution to the SR rate of a general $nlm$-state is unlikely to be very enlightening and in practice one would rather evaluate the full result numerically. 

To derive a simple analytic estimate, we focus instead on the leading $nlm=011$ state. In leading order in $\mu R_{\rm{LC}}$ we find the superradiance rate to be
\begin{align}
    \Gamma^{011}_{\text{SR}}&=\frac{\pi}{225}g_{\rm{a}\gamma\gamma}^2 B_0^2 (\mu R)^6 \mathcal{N}_{\phi}^2\alpha_{01}^5 \frac{\sigma}{\mu}(\mu R_{\rm{LC}})^5(\Omega- \frac{17}{15}\mu), \label{SRleading}
\end{align}
where $\omega\approx\mu$ has been used and $R_{\rm{LC}}\equiv\Omega^{-1}$ is the light cylinder radius of the star where magnetic field lines can no longer rigidly rotate with the star. In the radial integral we model the conductivity as in~\cite{Day_2019}, i.e. assume that $\sigma=$const. for $R\le r\le R_{\rm{LC}}$, $\sigma = \infty$ for $r<R$ and $\sigma=0$ for $r>R_{\rm LC}$.

\begin{figure}[h!tb]
    \centering
    \includegraphics[width=0.85\textwidth]{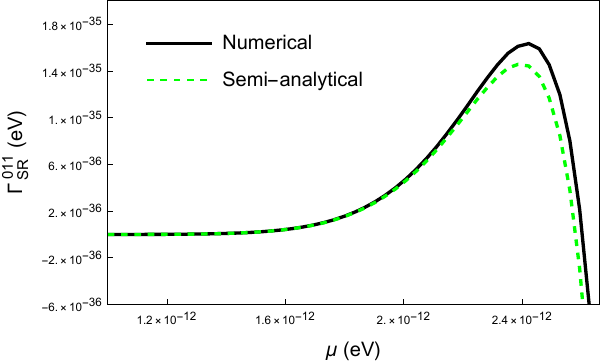}
    \caption{Comparison of the numerical result for the leading order superradiant rate and the analytic approximation in Eq.~\eqref{SRleading}. The parameters corresponding to PSR J0952-0607 are $M=2.35 M_{\odot} $, $R=12.5$ km, $B_0=10^8$ G, $\sigma = 10^{-12}$ eV and $\Omega = 2.92\times 10^{-12}$ eV. We have assumed the axion-photon coupling $g_{\rm{a}\gamma\gamma}=10^{-13}\ {\rm GeV}^{-1}$.}
    \label{fig:numvsanalytic}
\end{figure}

In Fig.~\ref{fig:numvsanalytic} we compare the leading-order semi-analytical rate of Eq.~\eqref{SRleading} and the  numerical result, obtained by evaluating Eq. \eqref{deltawnlm} using the matrix elements given in Eq. (\ref{phiVayA}) and in Appendix \ref{app:matrixes}. We find that they agree remarkably well. For higher axion masses the relation slowly fails as the leading order $\alpha\propto\mu$ expansion of the radial integrals ceases to be accurate enough\footnote{The gravitational coupling $\alpha$ is also proportional to the mass of the star. We have considered the most massive MSP known, PSR J0952-0607, for which $\alpha$ is maximal, to highlight that the semi-analytical formula is always valid.}. We note that the superradiance rate scales as $(\mu R)^6\alpha_{nl}^{2l+3}$. This has an extra factor of $\alpha_{nl}$ in relation to the result discussed in \cite{Day_2019}.

 We now outline the derivation of the semi-analytical expression of the superradiance rate for the $nlm=011$ state. To make the notation more convenient, we define first the quantities 
\begin{align}
    \kappa&\equiv g_{\rm{a}\gamma\gamma}B_0R^3\mathcal{N}_{\phi}\mu^3\alpha_{01}^{\frac{3}{2}}, \notag \\
    \gamma^{ll'}& \equiv \frac{\sigma}{\mu}\frac{(\mu R)^{l+l'+3}-(\mu R_{\rm{LC}})^{l+l'+3}}{2^{l+l'+1}\Gamma(l+\frac{3}{2})\Gamma(l'+\frac{3}{2})(l+l'+3)} \label{kappagamma}.
\end{align}
Furthermore, we define 
\begin{align} 
\langle \lambda \lambda' \rangle\equiv \left\langle\phi_{011}\left|V_{a \gamma \gamma}\right| A_{l m}^{(\lambda)}\right\rangle \left\langle A_{l^{\prime} m^{\prime}}^{(\lambda')}\left|V_{a \gamma \gamma}\right|\phi_{011}\right\rangle.
\end{align} 
 The selection rules now dictate e.g.
\begin{align}
    \langle 11 \rangle&=\frac{3}{2}\kappa ^2 I[l-2,l,l]^2 \big( S[1,l,2,1,m,1]S[1,l',2,1,m',1]\big)\notag \\ &\simeq\frac{9}{50}\kappa^2 I[-1,1,1]^2\delta_{1l'}\delta_{0m'}\delta_{1l}\delta_{0m} \simeq \frac{9}{50}\kappa^2 \alpha_{01}^2\delta_{1l'}\delta_{0m'}\delta_{1l}\delta_{0m} \label{11},
\end{align}
where we have taken into account that while $S[1,l,2,1,m,1]$ is nonzero also for higher (odd) values of $l$, these terms are suppressed by additional powers of $\mu R_{\rm{LC}}$ in Eq.~\eqref{kappagamma}. In other words, we keep only the dominant terms in the sum over intermediate states in Eq.~\eqref{deltawnlm}. In the final equality we have expanded the radial integral in $\alpha$ as demonstrated in Appendix~\ref{app:radialexp}. 

Finally, we denote by $\Gamma_{\lambda\lambda'}$ the term which appears in the expression for SR rate, Eq.~\eqref{deltawnlm}, and corresponds to $\langle \lambda \lambda' \rangle$,
\begin{align} 
\Gamma_{\lambda\lambda'}=\langle A^{(\lambda)}|V_A|A^{(\lambda')}\rangle\langle \lambda\lambda'\rangle.
\end{align}
We thus multiply Eq.~\eqref{11} by the corresponding photon-photon matrix element
\begin{align}
    \left\langle A_{l m}^{(1)}\left|V_A\right| A_{l^{\prime} m^{\prime}}^{(1)}\right\rangle&=i\gamma^{ll'} \big[\delta_{m m^{\prime}} \delta_{l l^{\prime}} -\frac{\Omega}{\mu}m'A[l,m,l',m',f_1^{11} ]+c_{l'm'}A[l,m,l',m'+1,f_2^{11}]\notag \\ & +i\frac{\Omega}{\mu}l' A[l,m,l',m',f_3^{11} ]\big].
\end{align}
The expressions for the matrix elements $\langle A^{(\lambda)}|V_A|A^{(\lambda')}\rangle$ can be found in Appendix~\ref{app:matrixes}. The rate $\Gamma_{11}$ is then given by 
\begin{align}
    \Gamma_{11}&=\frac{9}{50}i\kappa^2 \alpha_{01}^2\delta_{1l'}\delta_{0m'}\delta_{1l}\delta_{0m} \gamma^{ll'} \big[ \delta_{ll'}-\frac{\Omega}{\mu}m'A[l,m,l',m',f_1^{11} ]\notag \\ &+i\frac{\Omega}{\mu}c_{l'm'}A[l,m,l',m'+1,f_2^{11} ]+i\frac{\Omega}{\mu}l'A[l,m,l',m',f_3^{11} ]\big]\equiv\frac{9}{50}i\kappa^2 \gamma \alpha_{01}^2,
\end{align}
where we have further defined the shorthand $\gamma \equiv\gamma^{11}$. Note that $\Gamma_{11}$ is imaginary and contributes to $\delta\omega_{nlm}$. 
Similarly, omitting all nonzero terms that do not contribute to the imaginary part of $\delta\omega_{nlm}$, one eventually finds for all the spatial parts
\begin{align}
\rm{Im}(\Gamma_{12})&=\rm{Im}(\Gamma_{21})=\rm{Im}(\Gamma_{13})=0,\quad  
\rm{Im}(\Gamma_{22})=\rm{Im}(\Gamma_{11})=\frac{9}{50}\kappa^2\gamma\alpha_{01}^2\notag \\ 
\rm{Im}(\Gamma_{31})&=\rm{Im}(\Gamma_{32})=-\frac{6}{50}\kappa^2\gamma\frac{\Omega}{\mu}\alpha_{01}^2 ,\quad
\rm{Im}(\Gamma_{33})=\frac{8}{50}\kappa^2\gamma(1-\frac{\Omega}{\mu})\alpha_{01}^2, \label{spatials}
\end{align}
where we have dropped subleading contributions. For the temporal parts we again expand the relevant radial integrals in $\alpha_{01}$, as done in Appendix~\ref{app:radialexp}. In the end, we find the leading-order contributions to be given by
\begin{align}
    \rm{Im}(\Gamma_{10})&=\rm{Im}(\Gamma_{20})=-\frac{1}{50}\kappa^2\gamma\alpha_{01}^2,\quad 
  \rm{Im}(\Gamma_{01})=-\frac{1}{250}\kappa^2\frac{\Omega}{\mu} \alpha_{01}^2,\notag \\ 
  \rm{Im}(\Gamma_{02})&=\frac{1}{250}\kappa^2\gamma\frac{\Omega}{\mu}\alpha_{01}^2 =-\rm{Im}(\Gamma_{01}),\quad 
  \rm{Im}(\Gamma_{30})=-\frac{2}{75}\kappa^2 \gamma\alpha_{01}^2 .\label{timelikes}
\end{align}
Substituting Eqs.~\eqref{spatials} and~\eqref{timelikes} into
\begin{align}
    \Gamma^{011}_{\text{SR}}&=\frac{\pi^2}{8\mu}\sum_{\lambda\lambda'}\rm{Im}(\Gamma_{\lambda\lambda'})
\end{align}
one obtains the leading-order result of Eq.~\eqref{SRleading}.

\subsection{Constraints}

Rotation-powered pulsars lose angular momentum by dipole radiation. Other mechanisms are also known, such as magnetic breaking~\cite{harding2013neutronstarzoo}, and for the fastest millisecond pulsars also gravitational waves are emitted via the r-mode instability~\cite{Andersson:2000mf}. The rate of angular momentum loss and its derivatives are known to a very high precision by pulsar timing arrays and can be found in pulsar catalogues such as that of Australia Telescope National Facility (ATNF)~\cite{Manchester_2005}. For purely rotation-powered pulsars the characteristic spindown-time is given by
\begin{align}
    \tau_S=\frac{\Omega}{2\dot{\Omega}} = -\frac{\dot{P}}{2P}, \label{lifetime}
\end{align}
where $\Omega$ is the angular velocity of the star. Typically radio pulsars have lifetimes $10^5-10^9$ years while X-ray pulsars have slightly longer characteristic lifetimes of $10^9-10^{11}$ years \cite{harding2013neutronstarzoo}. The characteristic timescale of the superradiant instability on the other hand is given by $\tau_I=\Gamma^{-1}_{\rm{SR}}$. By requiring that observed stars are not dominated in their dynamics by the instability, $\tau_I/\tau_S>1$, we can place constraints on the axion-photon coupling parameter space. This leverages pulsar timing arrays, and is identical to requiring that SR-induced spindown rate not be larger than the observed spindown rate. In other words, the SR rate has to be such that $|\dot{\Omega}_{\rm{SR}}|\simeq|\Gamma_{\rm{SR}}\Omega|<\dot{\Omega}_{\rm{obs}}$.

\begin{figure}[t]
    \centering
    \includegraphics[width=0.9\textwidth]{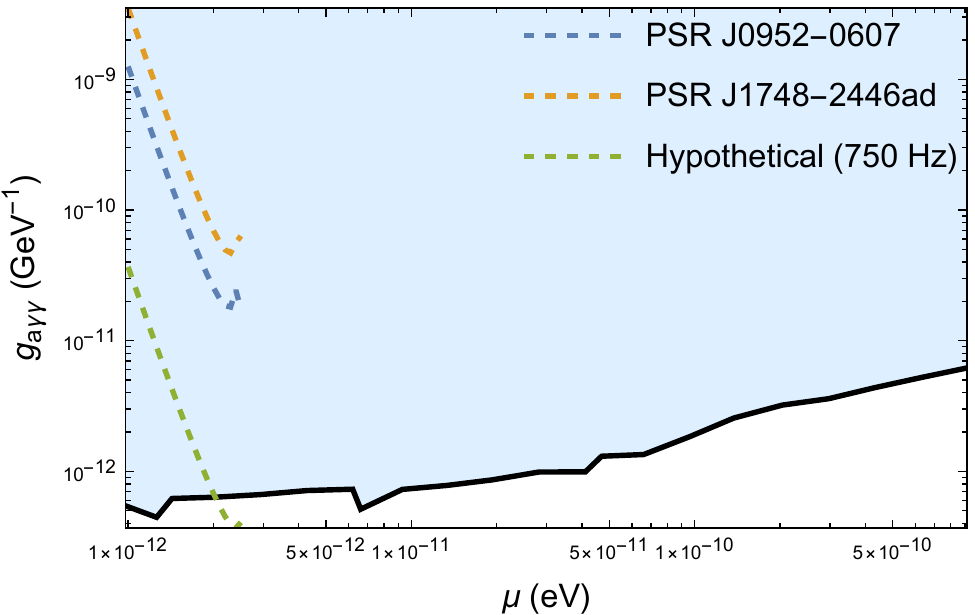}
    \caption{Projected bounds from the two largest spin MSPs known, as well as from a hypothetical 750 Hz pulsar with a stiff equation of state and assuming negligible plasma effects. The solid black line represents the envelope of currently leading constraints \cite{AxionLimits}.}
    \label{fig:constraints}
\end{figure}

The strongest constraints come from neutron stars which have long lifetimes and which have large masses and radii, as well as large magnetic fields. Of immediate interest are the fastest ($716$ Hz)  millisecond pulsar PSR J1748-2446ad considered in Ref.~\cite{Day_2019} as well as the more recently observed heaviest ($2.35 \ M_{\odot}$) and second fastest spinning ($707$ Hz) PSR J0952-0607~\cite{Romani_2022}. The mass and radius of PSR J1748-2446ad are not well known.  As MSPs typically form by accretion, we conservatively take them to be $M=1.4\ M_{\odot} $ and $R=10$ km. 
For PSR J0952-0607 the mass and radius are known better~\cite{Romani_2022}, and we take them to be $M=2.35 M_{\odot} $ and $R=12.5$ km. We assume the surface magnetic field to be $B_0=10^8$ G for both, and we take the conductivity in both cases to be $\sigma=10^{-12}$ eV, which is approximately in the middle of the observed range $0.01\Omega\lesssim \sigma \lesssim 100 \Omega$ ~\cite{Brambilla:2015vta} for pulsars. Finally, we take a hypothetical 750 Hz MSP with mass and radius again being given $M=2.35 M_{\odot} $ and $R=12.5$ km, but with slightly stronger magnetic field $5\times 10^8$ G. We take the Shklovskii-corrected period derivatives $\dot{P}$ needed to compute the characteristic lifetimes from the ATNF pulsar catalogue \cite{Manchester_2005}. For PSR J1748-2446ad, PSR J0952-0607 and the hypothetical pulsar we use $\dot{P}=7.05\times10^{-22}$, $\dot{P}=4.77\times10^{-21}$ and $\dot{P}=10^{-22}$ respectively. The projected bounds from each are shown in Fig.~\ref{fig:constraints} together with the current leading constraints from Ref.~\cite{AxionLimits}. The bounds are valid under the assumption that there is no strong suppression due to the photon plasma mass. The conditions under which this can be expected to hold are discussed in detail in Sec.~\ref{sec:plasmaeffects}.

As seen in Eq.~\eqref{SRleading}, there is a strong radius dependence in the superradiance rate, which means that stiffer neutron star equations of state lead to stronger constraints. This is clear from Fig.~\ref{fig:constraints} where the supermassive PSR J0952-0607 yields stronger bounds than the slightly faster PSR J1748-2446ad which was considered in~\cite{Day_2019}. Some hybrid equations of state could allow for more massive high spin MSPs~\cite{Gartlein:2024cbj}, thus considerably favoring the scenario we have considered here. 

The superradiance condition is of the form $m\Omega\gtrsim \mu$, where $m$ the azimuthal mode number. Therefore, one might wish to extend the axion mass range for which this mechanism is sensitive to by considering higher $m$ modes. However, the superradiance rate is then heavily suppressed by larger powers of $({\mu R_{\rm{LC}}})^l$.

\section{
Photon plasma mass and magnetosphere model}\label{sec:plasmaeffects}

In a plasma background the photon obtains an effective mass $\omega_p$ which can be larger than the mass of an axion capable of stellar superradiance. In such case the axion-photon conversion probability $p_{a\rightarrow\gamma}=|\langle A|V_{a\gamma}|\phi\rangle|^2$ is suppressed by an additional factor of $\sim(\frac{\omega}{\omega_p})^4$. As argued in Ref.~\cite{Spieksma:2025sda}, this plasma mass can effectively block a superradiantly scattered photon from depositing its energy back into the axion sector. One might also worry that the photon plasma mass, related to the imaginary part of the conductivity, might affect the superradiance mechanism itself. In this section we will briefly address the possible suppressive effect of the photon plasma mass and argue that the imaginary part of the conductivity does not qualitatively change the superradiance picture. We also justify the use of a global dipole magnetic field in this work.

The photon plasma mass, assuming a standard Goldreich-Julian (GJ) model for the magnetosphere with $n_e=|\frac{\Omega\cdot B}{2\pi e}|$ and a dipole magnetic field~\eqref{eq:DipoleField}, is given by
\begin{align}
    \omega_p &=\sqrt{\frac{n_{\mathrm{e}} e^2}{m_{\mathrm{e}}}} =\sqrt{\frac{e \Omega}{2\pi}\frac{B_0}{m_e}\frac{R^3}{r^3}(3\cos^2{\theta}-1)}, \label{gjwp}
\end{align}
which for a typical millisecond pulsar is around $\sim 10^{-8} \sqrt{(3\cos^2{\theta}-1)} \mathrm{\, eV} $, reaching as low as $\sim 10^{-9}\mathrm{ \,eV} $ at $\theta=53^\circ$ declination. Thus naively one would indeed expect superradiance to be strongly damped. However, 
we will now argue that 
already within the GJ model it is possible to significantly alleviate this suppression. Additionally, 
the GJ model 
may not always be a suitable to model millisecond pulsar magnetospheres to begin with, and that in other magnetospheric models the expected plasma suppression may even be negligible. 

Assuming a power law scaling $B =  B_\ast(\frac{R_\ast}{r})^\gamma$, 
a typical radial integral in an axion-photon matrix element, see Eq.~\eqref{Eq:auxdefs}, scales as
\begin{align}
    I(\gamma)&=B_\ast(\mu R_\ast)^\gamma \alpha^{\gamma-3/2}\int dx\, e^{-x/2}x^{l-\gamma+2}\left(1+\frac{\omega_p}{\mu}\right)^{-\frac{3}{2}}L_n^{2l+1}(x)j_{l'}\left(\frac{(\omega+\omega_p) x}{\mu\alpha}\right), \label{radint}
\end{align}
where the subscript $\ast$ is to be understood as the corresponding quantity at the surface of the star (for $r<R_{\rm{LC}}$) or at the light cylinder (for $r>R_{\rm{LC}}$). As the $\gamma$-scaling is generally different in the two regimes, one should evaluate the integral in two pieces with the above in mind. For the dipole field in Eq.~\eqref{eq:DipoleField}, we have $\gamma=3$.
 
 Numerically evaluating the integral in Eq.~\eqref{radint} we find that for the parameters relevant for superradiance, the radial integral contribution outside the light cylinder is generally dominant by factors of $\sim 10$ already for a dipole magnetic field ansatz regardless of plasma. Additionally, in the GJ model, the magnetic field can no longer corotate with the star outside the light cylinder, forming instead a pulsar wind which scales as $B\sim 1/r$ (i.e. $\gamma=1$) at large distances~\cite{Kirk:2007tn}. This further enhances the radial integral outside. 
 
 Inside the pulsar wind the electron density is typically assumed to fall as $n_e\sim 1/r^2$, under the assumption of particle number conservation, spherical symmetry, no collimation of the wind and no acceleration of particles. However, particles can be further accelerated in the pulsar wind to Lorentz factors up to $\sim 10^5$ or higher~\cite{Cerutti:2020qav}. This acceleration occurs to distance scales much larger than those relevant for superradiance, with the Lorentz factor growing approximately linearly with distance~\cite{1969ApJ...158..727M}. As a consequence, 
 the plasma density outside can fall off much faster. These considerations leave plausible the scenario of pulsar wind superradiance. 

In the GJ model it is assumed that the whole magnetosphere is filled with plasma. However, for pulsars of periods shorter than a second and magnetic field strengths below $B<10^9$ G, pair-creation is not efficient enough to fill the whole magnetosphere~\cite{2002A&A...384..414P}. Instead, there can be large regions of low charge density inside the pulsar magnetosphere. This occurs for instance in the electrosphere solution where charges are fully separated to domes of electrons around the polar caps and an equatorial torus of positrons and protons~\cite{2002A&A...384..414P}. This model is more realistic than the naive force-free magnetosphere for millisecond pulsar magnetospheres, and is observed to arise naturally in simulations when pair production rate is low~\cite{Guepin:2019fjb}. The distributions of all charged particles are shown concisely in Fig. 1 of Ref.~\cite{Guepin:2019fjb}. We note here that around the equator where the axion cloud is localized, the number densities of electrons, positrons and protons are given approximately by \mbox{$n_{\mathrm{e^-}}\approx n_{\mathrm{e^+}}\approx 0$} and \mbox{$n_{\mathrm{p}}\sim10^{-2}n_{GJ}$}. Due to their large inertia, the protons do not contribute significantly to the photon plasma mass. Thus, we expect that neglecting the plasma mass is an excellent approximation in this scenario. Additionally, as again seen in Fig.~1 of Ref.~\cite{Guepin:2019fjb}, in the electrosphere solution, away from the poles, the magnetic field lines close even beyond the light cylinder and behave as a dipole field. No causality requirement is broken by this, as there is no plasma moving along the field lines. This supports the global use of the dipole magnetic field in this work. In reality the nature of the magnetosphere is presumably something between the plasma-filled and charge-separated extremes. Indeed, in a perfectly charge-separated magnetosphere there would be no spindown, describing a completely dead pulsar~\cite{Cerutti:2016ttn}, and as such some plasma leakage is required.

Finally we briefly consider the impact of the imaginary part of the conductivity, connected to the photon plasma mass by $\sigma_I=\omega_p^2/\omega$, on the superradiance mechanism. We have verified that numerically the eigenfrequency correction of Eq.~\eqref{deltawnlm} is proportional to $i\sigma_R-\sigma_I$, so that the superradiance rate depends only on $\sigma_R$. This conclusion holds in the perturbative regime $\sigma_I,\sigma_R\ll \mu$, and for larger $\sigma_I$ one would expect some damping \cite{Cardoso_2017,Day_2019}. As noted in Ref. \cite{Spieksma:2025sda}, in the strongly-collisional regime $\omega\ll\nu$, with $\nu$ the collision rate in the Drude model, $\sigma_I\ll\sigma_R$ generally holds. This is true for instance in accreting pulsars in low-mass X-ray binaries.

\newpage
\section{Discussion and conclusions}
\label{sec:checkout}

In this paper we have computed perturbatively the superradiance rate for a finite-conductivity instability in the plasma using a dipole magnetic field configuration. We have applied this to constrain the axion-photon coupling for axion mass $\sim10^{-12}$ eV, using the fastest known MSP PSR J1748-2446ad and the supermassive PSR J0952–0607. Additionally we have given a projection for a hypothetical 750 Hz pulsar with a stiff equation of state. We find that known millisecond pulsars are unlikely to provide competitive bounds on axions. The hypothetical example shows, however, that if new fast MSPs are found with future radio telescopes, the situation could change. It has been recently noted in Ref.~\cite{Spieksma:2025sda}, that the superradiance rate could be additionally suppressed by the photon plasma mass. We have discussed how the magnitude of this effect is strongly dependent on the detailed structure of the pulsar magnetosphere: for force-free case the suppression due to the plasma mass is significant, while for the charge-separated magnetosphere we find that the effect is small. The reality lies in between and remains to be determined more precisely.

If such fast spinning pulsars with frequencies in the kHz range turn out not to exist at all, this would require an explanation. Such absence  could not be attributed to centrifugal breakup, as this happens only at higher spins \cite{Haskell:2018nlh}. With accretion spin-up times being much shorter than the observed lifetimes of stars, it seems there must be a mechanism that limits the maximum spin of a pulsar. Possible mechanisms of stopping spin-up include \cite{Cikintoglu:2023drm} gravitational waves, spin extraction by electromagnetic winds and the change of momentum of inertia due to accretion. Stellar superradiance could also provide such spin-down mechanism for stars. It would also lead to clustering of neutron star spins at the end point of the instability, $\Omega\sim\mu$, which could be observable in the pulsar population.

If a superradiant cloud does form around a neutron star by plasma or other instability, one might hope to be able observe its effect on the propagation of light. While the ultra-low energy photons from axion decay themselves are not observable, the polarization effects caused by the cloud may be observable in pulsar polarization arrays \cite{Liu:2021zlt}. It would also be interesting to consider the interaction between the other instabilities in neutron stars such as the r-mode instability \cite{Andersson:2000mf}. Neutron stars can also produce a superradiant cloud via dissipation provided by axion interactions with the stellar medium \cite{Chadha-Day:2022inf}. The interactions between these instabilities could be an interesting avenue to explore.

\begin{acknowledgments}
This work has been supported by the Research Council of Finland (grant\# 342777). TS would like to thank the Vilho, Yrjö and Kalle Väisälä Foundation. 
\end{acknowledgments}

\appendix

\section{Matrix elements}
\label{app:matrixes}

We list here all the photon-photon matrix elements required to reproduce the semi-analytical formula of Eq.~\eqref{SRleading} and to numerically evaluate the SR rate from Eq.~\eqref{deltawnlm}. In the radial integral we model the conductivity as in~\cite{Day_2019}, i.e. assume that $\sigma=$const. for $R\le r\le R_{\rm{LC}}=\Omega^{-1}$ and zero elsewhere. Here $R_{\rm{LC}}$ is the pulsar light-cylinder radius. We work in leading order in $\omega R\ll 1$.

We begin with the elements already determined in Ref. ~\cite{Day_2019}:
\begin{equation}
\langle A_{l m}^{(0)}\left|V_A\right| A_{l^{\prime} m^{\prime}}^{(0)}\rangle = i \sigma m \Omega \frac{\left( \omega R_{\mathrm{LC}}\right)^{2 l+3}-( \omega R)^{2 l+3}}{\omega^2 2^{2(l+1)}} \frac{1}{\left(l+\frac{3}{2}\right) \Gamma\left(l+\frac{3}{2}\right)^2} \delta_{m m^{\prime}} \delta_{l l^{\prime}},
\end{equation}
\begin{equation}
\langle A_{l m}^{(3)}\left|V_A\right| A_{l^{\prime} m^{\prime}}^{(3)}\rangle = i \sigma(m \Omega-\omega) \frac{\left( \omega R_{\mathrm{LC}}\right)^{2 l+3}-( \omega R)^{2 l+3}}{\omega^2 2^{2(l+1)}} \frac{1}{\left(l+\frac{3}{2}\right) \Gamma\left(l+\frac{3}{2}\right)^2} \delta_{m m^{\prime}} \delta_{l l^{\prime}},
\end{equation}
\begin{align}
\langle A_{l m}^{(3)}\left|V_A\right| A_{l^{\prime} m^{\prime}}^{(0)}\rangle &= \frac{\sigma }{\omega}\left[
\frac{(-1)^{-m}S\left[l, l',1,m,m',0 \right]\left[\left(\omega R_{L C}\right)^{l+l^{\prime}+2}-(\omega R)^{l+l^{\prime}+2}\right]}{2^{l+l'}\left(l+l'+2\right) \Gamma(l+\frac{3}{2})\Gamma(l'+\frac{1}{2})} \right. \notag \\ 
&\quad + \left. \frac{N_{l', m'}}{N_{l'+1, m'}} \delta_{l, l^{\prime}+1} \delta_{mm'} \frac{(m'-l'-1)\left[\left(\omega R\right)^{l+l'+2}-(\omega R_{L C})^{l+l'+2}\right]}{2^{l+l'} (2l'+3) \Gamma(l'+\frac{3}{2})\Gamma(l'+\frac{5}{2})} \right],
\end{align}
where $N_{lm}\equiv \sqrt{\frac{(2 l+1)(l-|m|)!}{4 \pi(l+|m|)!}}$. The matrix elements between the spatial polarizations are of form
\begin{align}
    \left\langle A_{l m}^{(i)}\left|V_A\right| A_{l^{\prime} m^{\prime}}^{(j)}\right\rangle&=i\gamma^{ll'} \big[\delta_{m m^{\prime}} \delta_{l l^{\prime}} -\frac{\Omega}{\mu}m'A[l,m,l',m',f^{ij}_1] \notag \\ &+c_{l'm'}A[l,m,l',m'+1,f^{ij}_2] +i\frac{\Omega}{\mu}l' A[l,m,l',m',f^{ij}_3]\big],
\end{align}
where the angular integrals $A[l,m,l',m',f]$ are given with respect to the auxiliary functions $f^{ij}_k$ with $k=1,2,3$, listed in Appendix \ref{app:functions}. The matrix elements mixing spatial and time polarizations are given by
\begin{align}
\langle A_{l m}^{(1)}|V_A| A_{l^{\prime} m^{\prime}}^{(0)}\rangle =& 
\frac{\sigma}{\mu} \frac{ (\omega R_{\text{LC}})^{l + l' + 2} - (\omega  R)^{l + l' + 2}}{2^{l + l' + 1} \Gamma\left(l + \frac{3}{2}\right) \Gamma\left(l' + \frac{3}{2}\right) (l + l' + 2)} 
( l' A[l,m,l',m',f_1^{10}]  \notag \\ +& m' A[l,m,l',m',f_2^{10}] + c_{lm} A[l,m,l',m'+1,f_3^{10}] ),
\notag \\
   \langle A_{l m}^{(2)}|V_A| A_{l^{\prime} m^{\prime}}^{(0)}\rangle =& 
\frac{\sigma}{\mu}  \frac{ (\omega  R_{\rm{LC}})^{l + l' + 2} - (\omega  R)^{l + l' + 2} }{2^{l + l' + 1} \Gamma\left(l + \frac{3}{2}\right) \Gamma\left(l' + \frac{3}{2}\right) (l + l' + 2)} 
( l' A[l,m,l',m',f_1^{20}] \notag \\ +& m' A[l,m,l',m',f_2^{20}] + c_{lm} A[l,m,l',m'+1,f_3^{20}] ),
\end{align}
\begin{align}
\langle A_{l m}^{(0)}\left|V_A\right| A_{l^{\prime} m^{\prime}}^{(1)}\rangle =&(-1)^{-m } \frac{\sigma \Omega}{\mu^2 \sqrt{2}} \frac{ (\omega  R_{\rm{LC}})^{l + l' + 4} - (\omega  R)^{l + l' + 4}}{2^{l + l' + 1} \Gamma\left(l + \frac{3}{2}\right) \Gamma\left(l' + \frac{3}{2}\right) (l + l' + 4)} \notag \\ &\times ( S[l,l',1,m,m',-1]+S[l,l',1,m,m',1] ), \notag \\
\langle A_{l m}^{(0)}\left|V_A\right| A_{l^{\prime} m^{\prime}}^{(2)}\rangle =i&(-1)^{-m+1 } \frac{\sigma \Omega}{\mu^2 \sqrt{2}} \ \frac{ (\omega  R_{\rm{LC}})^{l + l' + 4} - (\omega  R)^{l + l' + 4}}{2^{l + l' + 1} \Gamma\left(l + \frac{3}{2}\right) \Gamma\left(l' + \frac{3}{2}\right) (l + l' + 4)} \notag \\ &\times ( S[l,l',1,m,m',-1]-S[l,l',1,m,m',1] ),
\end{align}
\begin{equation}
\langle A_{l m}^{(0)}\left|V_A\right| A_{l^{\prime} m^{\prime}}^{(0)}\rangle =i \sigma m \Omega \frac{ (\omega  R_{\rm{LC}})^{2 l + 3} - (\omega  R)^{2 l + 3}}{2^{2 l + 1} (2 l + 3) \Gamma\left(l + \frac{3}{2}\right)^2 \mu^2} \, \delta_{l, l'} \delta_{m,m'},
\end{equation}
\begin{align}
\langle A_{l m}^{(3)}\left|V_A\right| A_{l^{\prime} m^{\prime}}^{(0)}\rangle =& 
\frac{\sigma}{\mu} \Big( \frac{ (\omega  R_{\rm{LC}})^{l + l' + 2} - (\omega  R)^{l + l' + 2} }{2^{l + l'} (l + l' + 2) \Gamma\left(l' + \frac{1}{2}\right) \Gamma\left(l + \frac{3}{2}\right)}(-1)^{-m} S[l,l',1,m,m',0]  \notag \\
&+\frac{N_{l',m'}}{N_{l'+1,m'}}(m'-l' - 1) \delta_{l, l' + 1} \delta_{m, m'} \frac{(\omega  R_{\rm{LC}})^{2 l' + 3} - (\omega  R)^{2 l' + 3} )}{2^{l' + 2} (2 l' + 3) \Gamma\left(l' + \frac{3}{2}\right) \Gamma\left(l' + \frac{5}{2}\right)} \Big).
\end{align}

\newpage 
\section{List of auxiliary functions}
\label{app:functions}

The angular integrals in the matrix elements are expressed in terms of angular integral functions $A[l_1,m_1,l_2,m_2,f^{ij}_k]$, as defined in Eq.~\eqref{Eq:auxdefs}. Here $f^{ij}_k$ with $k=1,2,3$ are particular combinations of trigonometric functions. They are:

\begin{alignat}{3}
    f_1^{11}  & = \ctsq\cph e^{-i\phi}+\stsq\cph^2 \qquad & f_2^{11}  & = \st\ct\sph\cph e^{-i\phi} \notag \\
    f_3^{11}  & = \stsq\sph\cph \qquad & f_1^{21}  & = \ctsq\sph e^{-i\phi}+\stsq\sph\cph \notag \\
    f_2^{21}  & = \st\ct\sphsq e^{-i\phi} \qquad & f_3^{21}  & = \stsq\sphsq \notag \\
    f_1^{31}  & = \st\ct\cph-\st\ct e^{-i\phi} \qquad & f_2^{31}  & = -\stsq\sph e^{-i\phi} \notag \\
    f_3^{31}  & = \st\ct\sph \qquad & f_{1}^{12} & = -\ctsq\cph e^{-i\phi}+i\stsq\sph\cph \notag \\
    f_{2}^{12} & = -\st\ct\cphsq e^{-i\phi} \qquad & f_{3}^{12} & = -\stsq\cphsq \notag \\
    f_{1}^{22} & = -\ctsq\sph e^{-i\phi}+i\stsq\sphsq \qquad & f_{2}^{22} & = -\st\ct\sph\cph e^{-i\phi} \notag \\
    f_{3}^{22} & = -\stsq\sph\cph \qquad & f_{1}^{32} & = \st\ct e^{-i\phi}+i\st\ct\sph \notag \\
    f_{2}^{32} & = \stsq\cph e^{-i\phi} \qquad & f_{3}^{32} & = -\st\ct\cph \notag \\
    f_{1}^{10} & = \st\cph \qquad & f_{2}^{10} & = \frac{\ctsq}{\st}\cph-i\frac{\sph}{\st} \notag \\
    f_{3}^{10} & = \ct\cph e^{-i\phi} \qquad & f_{1}^{20} & = \st \sph \notag \\
    f_{2}^{20} & = \frac{\ctsq}{\st}\sph+i\frac{\cph}{\st} \qquad & f_{3}^{20} & = \ct\sph e^{-i\phi} \label{functions}
\end{alignat}

\newpage
\section{Expansion of radial integrals}
\label{app:radialexp}

The radial integrals of form
\begin{align}
    I[l_1,l_2,l_3]\equiv & \int_0^\infty dx\ e^{\frac{x}{2}}x^{l_1+1}L_n^{2l_2+1}(x)j_{l_3}(\frac{x}{\alpha_{nl}})
\end{align}
can be expanded in the gravitational coupling $\alpha_{nl}\equiv\alpha$.
It can be shown, following similar steps to those of Appendix B in~\cite{Day_2019}, that the radial integrals of the form that appear in our calculation can generally be expanded in $\alpha$ as

\begin{align}
    I[l-a,l,l']\simeq & \pi{n+2l+1 \choose n}  2^{-l'-1}\alpha^{l+1-a}\Bigg( \frac{\alpha\Gamma(l+l'-a+2)}{\Gamma(\frac{1}{2}(-l+l'+a+1))\Gamma(\frac{1}{2}(l+l'-a+3))} \notag \\ &-\frac{\alpha^2\Gamma(l+l'-a+2)}{\Gamma(\frac{1}{2}(-l+l'+a))\Gamma(\frac{1}{2}(l+l'-a+2))}\Bigg). \label{radialexp}
\end{align}

In our case most of the radial integrals are of the form
\begin{align}
     I[l-2,l,l]\simeq 2^{l-1}\alpha^{l} {n+2l+1 \choose n}(l-1)!\xrightarrow{011}\alpha-\frac{\pi}{4}\alpha^2.
\end{align}

Additionally we need the integral
\begin{align}
     I[l-3,l,l]\simeq \frac{2^{l+1}}{3}\alpha^{l-1} {n+2l+1 \choose n}(l-1)!\xrightarrow{011}\frac{1}{3}-\frac{\pi}{8}\alpha
\end{align}
in the derivation of Eq.~\eqref{timelikes}.


\bibliography{draftrefs}

\begin{thebibliography}{64}%
\makeatletter
\providecommand \@ifxundefined [1]{%
 \@ifx{#1\undefined}
}%
\providecommand \@ifnum [1]{%
 \ifnum #1\expandafter \@firstoftwo
 \else \expandafter \@secondoftwo
 \fi
}%
\providecommand \@ifx [1]{%
 \ifx #1\expandafter \@firstoftwo
 \else \expandafter \@secondoftwo
 \fi
}%
\providecommand \natexlab [1]{#1}%
\providecommand \enquote  [1]{``#1''}%
\providecommand \bibnamefont  [1]{#1}%
\providecommand \bibfnamefont [1]{#1}%
\providecommand \citenamefont [1]{#1}%
\providecommand \href@noop [0]{\@secondoftwo}%
\providecommand \href [0]{\begingroup \@sanitize@url \@href}%
\providecommand \@href[1]{\@@startlink{#1}\@@href}%
\providecommand \@@href[1]{\endgroup#1\@@endlink}%
\providecommand \@sanitize@url [0]{\catcode `\\12\catcode `\$12\catcode
  `\&12\catcode `\#12\catcode `\^12\catcode `\_12\catcode `\%12\relax}%
\providecommand \@@startlink[1]{}%
\providecommand \@@endlink[0]{}%
\providecommand \url  [0]{\begingroup\@sanitize@url \@url }%
\providecommand \@url [1]{\endgroup\@href {#1}{\urlprefix }}%
\providecommand \urlprefix  [0]{URL }%
\providecommand \Eprint [0]{\href }%
\providecommand \doibase [0]{https://doi.org/}%
\providecommand \selectlanguage [0]{\@gobble}%
\providecommand \bibinfo  [0]{\@secondoftwo}%
\providecommand \bibfield  [0]{\@secondoftwo}%
\providecommand \translation [1]{[#1]}%
\providecommand \BibitemOpen [0]{}%
\providecommand \bibitemStop [0]{}%
\providecommand \bibitemNoStop [0]{.\EOS\space}%
\providecommand \EOS [0]{\spacefactor3000\relax}%
\providecommand \BibitemShut  [1]{\csname bibitem#1\endcsname}%
\let\auto@bib@innerbib\@empty
\bibitem [{\citenamefont {Peccei}\ and\ \citenamefont
  {Quinn}(1977)}]{Peccei:1977hh}%
  \BibitemOpen
  \bibfield  {author} {\bibinfo {author} {\bibfnamefont {R.~D.}\ \bibnamefont
  {Peccei}}\ and\ \bibinfo {author} {\bibfnamefont {H.~R.}\ \bibnamefont
  {Quinn}},\ }\bibfield  {title} {\bibinfo {title} {{CP Conservation in the
  Presence of Instantons}},\ }\href
  {https://doi.org/10.1103/PhysRevLett.38.1440} {\bibfield  {journal} {\bibinfo
   {journal} {Phys. Rev. Lett.}\ }\textbf {\bibinfo {volume} {38}},\ \bibinfo
  {pages} {1440} (\bibinfo {year} {1977})}\BibitemShut {NoStop}%
\bibitem [{\citenamefont {O'Hare}(2024)}]{OHare:2024nmr}%
  \BibitemOpen
  \bibfield  {author} {\bibinfo {author} {\bibfnamefont {C.~A.~J.}\
  \bibnamefont {O'Hare}},\ }\bibfield  {title} {\bibinfo {title} {{Cosmology of
  axion dark matter}},\ }\href {https://doi.org/10.22323/1.454.0040} {\bibfield
   {journal} {\bibinfo  {journal} {PoS}\ }\textbf {\bibinfo {volume}
  {COSMICWISPers}},\ \bibinfo {pages} {040} (\bibinfo {year} {2024})},\ \Eprint
  {https://arxiv.org/abs/2403.17697} {arXiv:2403.17697 [hep-ph]} \BibitemShut
  {NoStop}%
\bibitem [{\citenamefont {Marsh}(2016)}]{Marsh:2015xka}%
  \BibitemOpen
  \bibfield  {author} {\bibinfo {author} {\bibfnamefont {D.~J.~E.}\
  \bibnamefont {Marsh}},\ }\bibfield  {title} {\bibinfo {title} {{Axion
  Cosmology}},\ }\href {https://doi.org/10.1016/j.physrep.2016.06.005}
  {\bibfield  {journal} {\bibinfo  {journal} {Phys. Rept.}\ }\textbf {\bibinfo
  {volume} {643}},\ \bibinfo {pages} {1} (\bibinfo {year} {2016})},\ \Eprint
  {https://arxiv.org/abs/1510.07633} {arXiv:1510.07633 [astro-ph.CO]}
  \BibitemShut {NoStop}%
\bibitem [{\citenamefont {Hui}\ \emph {et~al.}(2017)\citenamefont {Hui},
  \citenamefont {Ostriker}, \citenamefont {Tremaine},\ and\ \citenamefont
  {Witten}}]{Hui:2016ltb}%
  \BibitemOpen
  \bibfield  {author} {\bibinfo {author} {\bibfnamefont {L.}~\bibnamefont
  {Hui}}, \bibinfo {author} {\bibfnamefont {J.~P.}\ \bibnamefont {Ostriker}},
  \bibinfo {author} {\bibfnamefont {S.}~\bibnamefont {Tremaine}},\ and\
  \bibinfo {author} {\bibfnamefont {E.}~\bibnamefont {Witten}},\ }\bibfield
  {title} {\bibinfo {title} {{Ultralight scalars as cosmological dark
  matter}},\ }\href {https://doi.org/10.1103/PhysRevD.95.043541} {\bibfield
  {journal} {\bibinfo  {journal} {Phys. Rev. D}\ }\textbf {\bibinfo {volume}
  {95}},\ \bibinfo {pages} {043541} (\bibinfo {year} {2017})},\ \Eprint
  {https://arxiv.org/abs/1610.08297} {arXiv:1610.08297 [astro-ph.CO]}
  \BibitemShut {NoStop}%
\bibitem [{\citenamefont {Di~Luzio}\ \emph {et~al.}(2020)\citenamefont
  {Di~Luzio}, \citenamefont {Giannotti}, \citenamefont {Nardi},\ and\
  \citenamefont {Visinelli}}]{DiLuzio:2020wdo}%
  \BibitemOpen
  \bibfield  {author} {\bibinfo {author} {\bibfnamefont {L.}~\bibnamefont
  {Di~Luzio}}, \bibinfo {author} {\bibfnamefont {M.}~\bibnamefont {Giannotti}},
  \bibinfo {author} {\bibfnamefont {E.}~\bibnamefont {Nardi}},\ and\ \bibinfo
  {author} {\bibfnamefont {L.}~\bibnamefont {Visinelli}},\ }\bibfield  {title}
  {\bibinfo {title} {{The landscape of QCD axion models}},\ }\href
  {https://doi.org/10.1016/j.physrep.2020.06.002} {\bibfield  {journal}
  {\bibinfo  {journal} {Phys. Rept.}\ }\textbf {\bibinfo {volume} {870}},\
  \bibinfo {pages} {1} (\bibinfo {year} {2020})},\ \Eprint
  {https://arxiv.org/abs/2003.01100} {arXiv:2003.01100 [hep-ph]} \BibitemShut
  {NoStop}%
\bibitem [{\citenamefont {Sikivie}(2021)}]{Sikivie:2020zpn}%
  \BibitemOpen
  \bibfield  {author} {\bibinfo {author} {\bibfnamefont {P.}~\bibnamefont
  {Sikivie}},\ }\bibfield  {title} {\bibinfo {title} {{Invisible Axion Search
  Methods}},\ }\href {https://doi.org/10.1103/RevModPhys.93.015004} {\bibfield
  {journal} {\bibinfo  {journal} {Rev. Mod. Phys.}\ }\textbf {\bibinfo {volume}
  {93}},\ \bibinfo {pages} {015004} (\bibinfo {year} {2021})},\ \Eprint
  {https://arxiv.org/abs/2003.02206} {arXiv:2003.02206 [hep-ph]} \BibitemShut
  {NoStop}%
\bibitem [{\citenamefont {Holdom}(1986)}]{Holdom:1985ag}%
  \BibitemOpen
  \bibfield  {author} {\bibinfo {author} {\bibfnamefont {B.}~\bibnamefont
  {Holdom}},\ }\bibfield  {title} {\bibinfo {title} {{Two U(1)'s and Epsilon
  Charge Shifts}},\ }\href {https://doi.org/10.1016/0370-2693(86)91377-8}
  {\bibfield  {journal} {\bibinfo  {journal} {Phys. Lett. B}\ }\textbf
  {\bibinfo {volume} {166}},\ \bibinfo {pages} {196} (\bibinfo {year}
  {1986})}\BibitemShut {NoStop}%
\bibitem [{\citenamefont {Fabbrichesi}\ \emph {et~al.}(2020)\citenamefont
  {Fabbrichesi}, \citenamefont {Gabrielli},\ and\ \citenamefont
  {Lanfranchi}}]{Fabbrichesi:2020wbt}%
  \BibitemOpen
  \bibfield  {author} {\bibinfo {author} {\bibfnamefont {M.}~\bibnamefont
  {Fabbrichesi}}, \bibinfo {author} {\bibfnamefont {E.}~\bibnamefont
  {Gabrielli}},\ and\ \bibinfo {author} {\bibfnamefont {G.}~\bibnamefont
  {Lanfranchi}},\ }\bibfield  {title} {\bibinfo {title} {{The Dark Photon}}\
  }\href {https://doi.org/10.1007/978-3-030-62519-1}
  {10.1007/978-3-030-62519-1} (\bibinfo {year} {2020}),\ \Eprint
  {https://arxiv.org/abs/2005.01515} {arXiv:2005.01515 [hep-ph]} \BibitemShut
  {NoStop}%
\bibitem [{\citenamefont {Weinberg}(1978)}]{Weinberg:1977ma}%
  \BibitemOpen
  \bibfield  {author} {\bibinfo {author} {\bibfnamefont {S.}~\bibnamefont
  {Weinberg}},\ }\bibfield  {title} {\bibinfo {title} {{A New Light Boson?}},\
  }\href {https://doi.org/10.1103/PhysRevLett.40.223} {\bibfield  {journal}
  {\bibinfo  {journal} {Phys. Rev. Lett.}\ }\textbf {\bibinfo {volume} {40}},\
  \bibinfo {pages} {223} (\bibinfo {year} {1978})}\BibitemShut {NoStop}%
\bibitem [{\citenamefont {O'Hare}(2020)}]{AxionLimits}%
  \BibitemOpen
  \bibfield  {author} {\bibinfo {author} {\bibfnamefont {C.}~\bibnamefont
  {O'Hare}},\ }\href {https://doi.org/10.5281/zenodo.3932430} {\bibinfo {title}
  {cajohare/axionlimits: Axionlimits}},\ \bibinfo {howpublished}
  {\url{https://cajohare.github.io/AxionLimits/}} (\bibinfo {year}
  {2020})\BibitemShut {NoStop}%
\bibitem [{\citenamefont {Xue}\ \emph {et~al.}(2024)\citenamefont {Xue} \emph
  {et~al.}}]{Xue:2024zjq}%
  \BibitemOpen
  \bibfield  {author} {\bibinfo {author} {\bibfnamefont {X.}~\bibnamefont
  {Xue}} \emph {et~al.},\ }\bibfield  {title} {\bibinfo {title} {{First Pulsar
  Polarization Array Limits on Ultralight Axion-like Dark Matter}},\
  }\href@noop {} {\  (\bibinfo {year} {2024})},\ \Eprint
  {https://arxiv.org/abs/2412.02229} {arXiv:2412.02229 [astro-ph.HE]}
  \BibitemShut {NoStop}%
\bibitem [{\citenamefont {Fedderke}\ \emph {et~al.}(2019)\citenamefont
  {Fedderke}, \citenamefont {Graham},\ and\ \citenamefont
  {Rajendran}}]{Fedderke:2019ajk}%
  \BibitemOpen
  \bibfield  {author} {\bibinfo {author} {\bibfnamefont {M.~A.}\ \bibnamefont
  {Fedderke}}, \bibinfo {author} {\bibfnamefont {P.~W.}\ \bibnamefont
  {Graham}},\ and\ \bibinfo {author} {\bibfnamefont {S.}~\bibnamefont
  {Rajendran}},\ }\bibfield  {title} {\bibinfo {title} {{Axion Dark Matter
  Detection with CMB Polarization}},\ }\href
  {https://doi.org/10.1103/PhysRevD.100.015040} {\bibfield  {journal} {\bibinfo
   {journal} {Phys. Rev. D}\ }\textbf {\bibinfo {volume} {100}},\ \bibinfo
  {pages} {015040} (\bibinfo {year} {2019})},\ \Eprint
  {https://arxiv.org/abs/1903.02666} {arXiv:1903.02666 [astro-ph.CO]}
  \BibitemShut {NoStop}%
\bibitem [{\citenamefont {Gan}\ \emph {et~al.}(2024)\citenamefont {Gan},
  \citenamefont {Wang},\ and\ \citenamefont {Xiao}}]{Gan:2023swl}%
  \BibitemOpen
  \bibfield  {author} {\bibinfo {author} {\bibfnamefont {X.}~\bibnamefont
  {Gan}}, \bibinfo {author} {\bibfnamefont {L.-T.}\ \bibnamefont {Wang}},\ and\
  \bibinfo {author} {\bibfnamefont {H.}~\bibnamefont {Xiao}},\ }\bibfield
  {title} {\bibinfo {title} {{Detecting axion dark matter with black hole
  polarimetry}},\ }\href {https://doi.org/10.1103/PhysRevD.110.063039}
  {\bibfield  {journal} {\bibinfo  {journal} {Phys. Rev. D}\ }\textbf {\bibinfo
  {volume} {110}},\ \bibinfo {pages} {063039} (\bibinfo {year} {2024})},\
  \Eprint {https://arxiv.org/abs/2311.02149} {arXiv:2311.02149 [hep-ph]}
  \BibitemShut {NoStop}%
\bibitem [{\citenamefont {Goodman}\ \emph {et~al.}(2025)\citenamefont {Goodman}
  \emph {et~al.}}]{ADMX:2024xbv}%
  \BibitemOpen
  \bibfield  {author} {\bibinfo {author} {\bibfnamefont {C.}~\bibnamefont
  {Goodman}} \emph {et~al.} (\bibinfo {collaboration} {ADMX}),\ }\bibfield
  {title} {\bibinfo {title} {{ADMX Axion Dark Matter Bounds around
  3.3\,\,\ensuremath{\mu}eV with Dine-Fischler-Srednicki-Zhitnitsky Discovery
  Ability}},\ }\href {https://doi.org/10.1103/PhysRevLett.134.111002}
  {\bibfield  {journal} {\bibinfo  {journal} {Phys. Rev. Lett.}\ }\textbf
  {\bibinfo {volume} {134}},\ \bibinfo {pages} {111002} (\bibinfo {year}
  {2025})},\ \Eprint {https://arxiv.org/abs/2408.15227} {arXiv:2408.15227
  [hep-ex]} \BibitemShut {NoStop}%
\bibitem [{\citenamefont {Adair}\ \emph {et~al.}(2022)\citenamefont {Adair}
  \emph {et~al.}}]{Adair:2022rtw}%
  \BibitemOpen
  \bibfield  {author} {\bibinfo {author} {\bibfnamefont {C.~M.}\ \bibnamefont
  {Adair}} \emph {et~al.},\ }\bibfield  {title} {\bibinfo {title} {{Search for
  Dark Matter Axions with CAST-CAPP}},\ }\href
  {https://doi.org/10.1038/s41467-022-33913-6} {\bibfield  {journal} {\bibinfo
  {journal} {Nature Commun.}\ }\textbf {\bibinfo {volume} {13}},\ \bibinfo
  {pages} {6180} (\bibinfo {year} {2022})},\ \Eprint
  {https://arxiv.org/abs/2211.02902} {arXiv:2211.02902 [hep-ex]} \BibitemShut
  {NoStop}%
\bibitem [{\citenamefont {Anastassopoulos}\ \emph {et~al.}(2017)\citenamefont
  {Anastassopoulos} \emph {et~al.}}]{CAST:2017uph}%
  \BibitemOpen
  \bibfield  {author} {\bibinfo {author} {\bibfnamefont {V.}~\bibnamefont
  {Anastassopoulos}} \emph {et~al.} (\bibinfo {collaboration} {CAST}),\
  }\bibfield  {title} {\bibinfo {title} {{New CAST Limit on the Axion-Photon
  Interaction}},\ }\href {https://doi.org/10.1038/nphys4109} {\bibfield
  {journal} {\bibinfo  {journal} {Nature Phys.}\ }\textbf {\bibinfo {volume}
  {13}},\ \bibinfo {pages} {584} (\bibinfo {year} {2017})},\ \Eprint
  {https://arxiv.org/abs/1705.02290} {arXiv:1705.02290 [hep-ex]} \BibitemShut
  {NoStop}%
\bibitem [{\citenamefont {Reyn\'es}\ \emph {et~al.}(2021)\citenamefont
  {Reyn\'es}, \citenamefont {Matthews}, \citenamefont {Reynolds}, \citenamefont
  {Russell}, \citenamefont {Smith},\ and\ \citenamefont
  {Marsh}}]{Reynes:2021bpe}%
  \BibitemOpen
  \bibfield  {author} {\bibinfo {author} {\bibfnamefont {J.~S.}\ \bibnamefont
  {Reyn\'es}}, \bibinfo {author} {\bibfnamefont {J.~H.}\ \bibnamefont
  {Matthews}}, \bibinfo {author} {\bibfnamefont {C.~S.}\ \bibnamefont
  {Reynolds}}, \bibinfo {author} {\bibfnamefont {H.~R.}\ \bibnamefont
  {Russell}}, \bibinfo {author} {\bibfnamefont {R.~N.}\ \bibnamefont {Smith}},\
  and\ \bibinfo {author} {\bibfnamefont {M.~C.~D.}\ \bibnamefont {Marsh}},\
  }\bibfield  {title} {\bibinfo {title} {{New constraints on light axion-like
  particles using Chandra transmission grating spectroscopy of the powerful
  cluster-hosted quasar H1821+643}},\ }\href
  {https://doi.org/10.1093/mnras/stab3464} {\bibfield  {journal} {\bibinfo
  {journal} {Mon. Not. Roy. Astron. Soc.}\ }\textbf {\bibinfo {volume} {510}},\
  \bibinfo {pages} {1264} (\bibinfo {year} {2021})},\ \Eprint
  {https://arxiv.org/abs/2109.03261} {arXiv:2109.03261 [astro-ph.HE]}
  \BibitemShut {NoStop}%
\bibitem [{\citenamefont {Ning}\ and\ \citenamefont
  {Safdi}(2024)}]{Ning:2024eky}%
  \BibitemOpen
  \bibfield  {author} {\bibinfo {author} {\bibfnamefont {O.}~\bibnamefont
  {Ning}}\ and\ \bibinfo {author} {\bibfnamefont {B.~R.}\ \bibnamefont
  {Safdi}},\ }\bibfield  {title} {\bibinfo {title} {{Leading Axion-Photon
  Sensitivity with NuSTAR Observations of M82 and M87}},\ }\href@noop {} {\
  (\bibinfo {year} {2024})},\ \Eprint {https://arxiv.org/abs/2404.14476}
  {arXiv:2404.14476 [hep-ph]} \BibitemShut {NoStop}%
\bibitem [{\citenamefont {Leroy}\ \emph {et~al.}(2020)\citenamefont {Leroy},
  \citenamefont {Chianese}, \citenamefont {Edwards},\ and\ \citenamefont
  {Weniger}}]{Leroy:2019ghm}%
  \BibitemOpen
  \bibfield  {author} {\bibinfo {author} {\bibfnamefont {M.}~\bibnamefont
  {Leroy}}, \bibinfo {author} {\bibfnamefont {M.}~\bibnamefont {Chianese}},
  \bibinfo {author} {\bibfnamefont {T.~D.~P.}\ \bibnamefont {Edwards}},\ and\
  \bibinfo {author} {\bibfnamefont {C.}~\bibnamefont {Weniger}},\ }\bibfield
  {title} {\bibinfo {title} {{Radio Signal of Axion-Photon Conversion in
  Neutron Stars: A Ray Tracing Analysis}},\ }\href
  {https://doi.org/10.1103/PhysRevD.101.123003} {\bibfield  {journal} {\bibinfo
   {journal} {Phys. Rev. D}\ }\textbf {\bibinfo {volume} {101}},\ \bibinfo
  {pages} {123003} (\bibinfo {year} {2020})},\ \Eprint
  {https://arxiv.org/abs/1912.08815} {arXiv:1912.08815 [hep-ph]} \BibitemShut
  {NoStop}%
\bibitem [{\citenamefont {Battye}\ \emph {et~al.}(2023)\citenamefont {Battye},
  \citenamefont {Keith}, \citenamefont {McDonald}, \citenamefont {Srinivasan},
  \citenamefont {Stappers},\ and\ \citenamefont {Weltevrede}}]{Battye:2023oac}%
  \BibitemOpen
  \bibfield  {author} {\bibinfo {author} {\bibfnamefont {R.~A.}\ \bibnamefont
  {Battye}}, \bibinfo {author} {\bibfnamefont {M.~J.}\ \bibnamefont {Keith}},
  \bibinfo {author} {\bibfnamefont {J.~I.}\ \bibnamefont {McDonald}}, \bibinfo
  {author} {\bibfnamefont {S.}~\bibnamefont {Srinivasan}}, \bibinfo {author}
  {\bibfnamefont {B.~W.}\ \bibnamefont {Stappers}},\ and\ \bibinfo {author}
  {\bibfnamefont {P.}~\bibnamefont {Weltevrede}},\ }\bibfield  {title}
  {\bibinfo {title} {{Searching for time-dependent axion dark matter signals in
  pulsars}},\ }\href {https://doi.org/10.1103/PhysRevD.108.063001} {\bibfield
  {journal} {\bibinfo  {journal} {Phys. Rev. D}\ }\textbf {\bibinfo {volume}
  {108}},\ \bibinfo {pages} {063001} (\bibinfo {year} {2023})},\ \Eprint
  {https://arxiv.org/abs/2303.11792} {arXiv:2303.11792 [astro-ph.CO]}
  \BibitemShut {NoStop}%
\bibitem [{\citenamefont {Prabhu}(2021)}]{Prabhu:2021zve}%
  \BibitemOpen
  \bibfield  {author} {\bibinfo {author} {\bibfnamefont {A.}~\bibnamefont
  {Prabhu}},\ }\bibfield  {title} {\bibinfo {title} {{Axion production in
  pulsar magnetosphere gaps}},\ }\href
  {https://doi.org/10.1103/PhysRevD.104.055038} {\bibfield  {journal} {\bibinfo
   {journal} {Phys. Rev. D}\ }\textbf {\bibinfo {volume} {104}},\ \bibinfo
  {pages} {055038} (\bibinfo {year} {2021})},\ \Eprint
  {https://arxiv.org/abs/2104.14569} {arXiv:2104.14569 [hep-ph]} \BibitemShut
  {NoStop}%
\bibitem [{\citenamefont {Witte}\ \emph {et~al.}(2021)\citenamefont {Witte},
  \citenamefont {Noordhuis}, \citenamefont {Edwards},\ and\ \citenamefont
  {Weniger}}]{Witte:2021arp}%
  \BibitemOpen
  \bibfield  {author} {\bibinfo {author} {\bibfnamefont {S.~J.}\ \bibnamefont
  {Witte}}, \bibinfo {author} {\bibfnamefont {D.}~\bibnamefont {Noordhuis}},
  \bibinfo {author} {\bibfnamefont {T.~D.~P.}\ \bibnamefont {Edwards}},\ and\
  \bibinfo {author} {\bibfnamefont {C.}~\bibnamefont {Weniger}},\ }\bibfield
  {title} {\bibinfo {title} {{Axion-photon conversion in neutron star
  magnetospheres: The role of the plasma in the Goldreich-Julian model}},\
  }\href {https://doi.org/10.1103/PhysRevD.104.103030} {\bibfield  {journal}
  {\bibinfo  {journal} {Phys. Rev. D}\ }\textbf {\bibinfo {volume} {104}},\
  \bibinfo {pages} {103030} (\bibinfo {year} {2021})},\ \Eprint
  {https://arxiv.org/abs/2104.07670} {arXiv:2104.07670 [hep-ph]} \BibitemShut
  {NoStop}%
\bibitem [{\citenamefont {Witte}\ \emph {et~al.}(2024)\citenamefont {Witte},
  \citenamefont {Noordhuis}, \citenamefont {Prabhu},\ and\ \citenamefont
  {Weniger}}]{Witte:2024akb}%
  \BibitemOpen
  \bibfield  {author} {\bibinfo {author} {\bibfnamefont {S.}~\bibnamefont
  {Witte}}, \bibinfo {author} {\bibfnamefont {D.}~\bibnamefont {Noordhuis}},
  \bibinfo {author} {\bibfnamefont {A.}~\bibnamefont {Prabhu}},\ and\ \bibinfo
  {author} {\bibfnamefont {C.}~\bibnamefont {Weniger}},\ }\bibfield  {title}
  {\bibinfo {title} {{The Growth and Evolution of Axion Clouds around
  Pulsars}},\ }\href {https://doi.org/10.22323/1.454.0022} {\bibfield
  {journal} {\bibinfo  {journal} {PoS}\ }\textbf {\bibinfo {volume}
  {COSMICWISPers}},\ \bibinfo {pages} {022} (\bibinfo {year}
  {2024})}\BibitemShut {NoStop}%
\bibitem [{\citenamefont {Noordhuis}\ \emph {et~al.}(2024)\citenamefont
  {Noordhuis}, \citenamefont {Prabhu}, \citenamefont {Weniger},\ and\
  \citenamefont {Witte}}]{Noordhuis:2023wid}%
  \BibitemOpen
  \bibfield  {author} {\bibinfo {author} {\bibfnamefont {D.}~\bibnamefont
  {Noordhuis}}, \bibinfo {author} {\bibfnamefont {A.}~\bibnamefont {Prabhu}},
  \bibinfo {author} {\bibfnamefont {C.}~\bibnamefont {Weniger}},\ and\ \bibinfo
  {author} {\bibfnamefont {S.~J.}\ \bibnamefont {Witte}},\ }\bibfield  {title}
  {\bibinfo {title} {{Axion Clouds around Neutron Stars}},\ }\href
  {https://doi.org/10.1103/PhysRevX.14.041015} {\bibfield  {journal} {\bibinfo
  {journal} {Phys. Rev. X}\ }\textbf {\bibinfo {volume} {14}},\ \bibinfo
  {pages} {041015} (\bibinfo {year} {2024})},\ \Eprint
  {https://arxiv.org/abs/2307.11811} {arXiv:2307.11811 [hep-ph]} \BibitemShut
  {NoStop}%
\bibitem [{\citenamefont {Detweiler}(1980)}]{PhysRevD.22.2323}%
  \BibitemOpen
  \bibfield  {author} {\bibinfo {author} {\bibfnamefont {S.}~\bibnamefont
  {Detweiler}},\ }\bibfield  {title} {\bibinfo {title} {Klein-gordon equation
  and rotating black holes},\ }\href {https://doi.org/10.1103/PhysRevD.22.2323}
  {\bibfield  {journal} {\bibinfo  {journal} {Phys. Rev. D}\ }\textbf {\bibinfo
  {volume} {22}},\ \bibinfo {pages} {2323} (\bibinfo {year}
  {1980})}\BibitemShut {NoStop}%
\bibitem [{\citenamefont {Arvanitaki}\ \emph {et~al.}(2010)\citenamefont
  {Arvanitaki}, \citenamefont {Dimopoulos}, \citenamefont {Dubovsky},
  \citenamefont {Kaloper},\ and\ \citenamefont
  {March-Russell}}]{Arvanitaki:2009fg}%
  \BibitemOpen
  \bibfield  {author} {\bibinfo {author} {\bibfnamefont {A.}~\bibnamefont
  {Arvanitaki}}, \bibinfo {author} {\bibfnamefont {S.}~\bibnamefont
  {Dimopoulos}}, \bibinfo {author} {\bibfnamefont {S.}~\bibnamefont
  {Dubovsky}}, \bibinfo {author} {\bibfnamefont {N.}~\bibnamefont {Kaloper}},\
  and\ \bibinfo {author} {\bibfnamefont {J.}~\bibnamefont {March-Russell}},\
  }\bibfield  {title} {\bibinfo {title} {{String Axiverse}},\ }\href
  {https://doi.org/10.1103/PhysRevD.81.123530} {\bibfield  {journal} {\bibinfo
  {journal} {Phys. Rev. D}\ }\textbf {\bibinfo {volume} {81}},\ \bibinfo
  {pages} {123530} (\bibinfo {year} {2010})},\ \Eprint
  {https://arxiv.org/abs/0905.4720} {arXiv:0905.4720 [hep-th]} \BibitemShut
  {NoStop}%
\bibitem [{\citenamefont {Witte}\ and\ \citenamefont
  {Mummery}(2024)}]{Witte:2024drg}%
  \BibitemOpen
  \bibfield  {author} {\bibinfo {author} {\bibfnamefont {S.~J.}\ \bibnamefont
  {Witte}}\ and\ \bibinfo {author} {\bibfnamefont {A.}~\bibnamefont
  {Mummery}},\ }\bibfield  {title} {\bibinfo {title} {{Stepping Up
  Superradiance Constraints on Axions}},\ }\href@noop {} {\  (\bibinfo {year}
  {2024})},\ \Eprint {https://arxiv.org/abs/2412.03655} {arXiv:2412.03655
  [hep-ph]} \BibitemShut {NoStop}%
\bibitem [{\citenamefont {Baryakhtar}\ \emph {et~al.}(2021)\citenamefont
  {Baryakhtar}, \citenamefont {Galanis}, \citenamefont {Lasenby},\ and\
  \citenamefont {Simon}}]{Baryakhtar:2020gao}%
  \BibitemOpen
  \bibfield  {author} {\bibinfo {author} {\bibfnamefont {M.}~\bibnamefont
  {Baryakhtar}}, \bibinfo {author} {\bibfnamefont {M.}~\bibnamefont {Galanis}},
  \bibinfo {author} {\bibfnamefont {R.}~\bibnamefont {Lasenby}},\ and\ \bibinfo
  {author} {\bibfnamefont {O.}~\bibnamefont {Simon}},\ }\bibfield  {title}
  {\bibinfo {title} {{Black hole superradiance of self-interacting scalar
  fields}},\ }\href {https://doi.org/10.1103/PhysRevD.103.095019} {\bibfield
  {journal} {\bibinfo  {journal} {Phys. Rev. D}\ }\textbf {\bibinfo {volume}
  {103}},\ \bibinfo {pages} {095019} (\bibinfo {year} {2021})},\ \Eprint
  {https://arxiv.org/abs/2011.11646} {arXiv:2011.11646 [hep-ph]} \BibitemShut
  {NoStop}%
\bibitem [{\citenamefont {Stott}(2020)}]{Stott:2020gjj}%
  \BibitemOpen
  \bibfield  {author} {\bibinfo {author} {\bibfnamefont {M.~J.}\ \bibnamefont
  {Stott}},\ }\bibfield  {title} {\bibinfo {title} {{Ultralight Bosonic Field
  Mass Bounds from Astrophysical Black Hole Spin}},\ }\href@noop {} {\
  (\bibinfo {year} {2020})},\ \Eprint {https://arxiv.org/abs/2009.07206}
  {arXiv:2009.07206 [hep-ph]} \BibitemShut {NoStop}%
\bibitem [{\citenamefont {Hoof}\ \emph {et~al.}(2024)\citenamefont {Hoof},
  \citenamefont {Marsh}, \citenamefont {Sisk-Reyn\'es}, \citenamefont
  {Matthews},\ and\ \citenamefont {Reynolds}}]{Hoof:2024quk}%
  \BibitemOpen
  \bibfield  {author} {\bibinfo {author} {\bibfnamefont {S.}~\bibnamefont
  {Hoof}}, \bibinfo {author} {\bibfnamefont {D.~J.~E.}\ \bibnamefont {Marsh}},
  \bibinfo {author} {\bibfnamefont {J.}~\bibnamefont {Sisk-Reyn\'es}}, \bibinfo
  {author} {\bibfnamefont {J.~H.}\ \bibnamefont {Matthews}},\ and\ \bibinfo
  {author} {\bibfnamefont {C.}~\bibnamefont {Reynolds}},\ }\bibfield  {title}
  {\bibinfo {title} {{Getting More Out of Black Hole Superradiance: a
  Statistically Rigorous Approach to Ultralight Boson Constraints}},\
  }\href@noop {} {\  (\bibinfo {year} {2024})},\ \Eprint
  {https://arxiv.org/abs/2406.10337} {arXiv:2406.10337 [hep-ph]} \BibitemShut
  {NoStop}%
\bibitem [{\citenamefont {\"Unal}\ \emph {et~al.}(2021)\citenamefont {\"Unal},
  \citenamefont {Pacucci},\ and\ \citenamefont {Loeb}}]{Unal:2020jiy}%
  \BibitemOpen
  \bibfield  {author} {\bibinfo {author} {\bibfnamefont {C.}~\bibnamefont
  {\"Unal}}, \bibinfo {author} {\bibfnamefont {F.}~\bibnamefont {Pacucci}},\
  and\ \bibinfo {author} {\bibfnamefont {A.}~\bibnamefont {Loeb}},\ }\bibfield
  {title} {\bibinfo {title} {{Properties of ultralight bosons from heavy quasar
  spins via superradiance}},\ }\href
  {https://doi.org/10.1088/1475-7516/2021/05/007} {\bibfield  {journal}
  {\bibinfo  {journal} {JCAP}\ }\textbf {\bibinfo {volume} {05}},\ \bibinfo
  {pages} {007}},\ \Eprint {https://arxiv.org/abs/2012.12790} {arXiv:2012.12790
  [hep-ph]} \BibitemShut {NoStop}%
\bibitem [{\citenamefont {Omiya}\ \emph {et~al.}(2024)\citenamefont {Omiya},
  \citenamefont {Takahashi}, \citenamefont {Tanaka},\ and\ \citenamefont
  {Yoshino}}]{Omiya:2024xlz}%
  \BibitemOpen
  \bibfield  {author} {\bibinfo {author} {\bibfnamefont {H.}~\bibnamefont
  {Omiya}}, \bibinfo {author} {\bibfnamefont {T.}~\bibnamefont {Takahashi}},
  \bibinfo {author} {\bibfnamefont {T.}~\bibnamefont {Tanaka}},\ and\ \bibinfo
  {author} {\bibfnamefont {H.}~\bibnamefont {Yoshino}},\ }\bibfield  {title}
  {\bibinfo {title} {{Deci-Hz gravitational waves from the self-interacting
  axion cloud around a rotating stellar mass black hole}},\ }\href
  {https://doi.org/10.1103/PhysRevD.110.044002} {\bibfield  {journal} {\bibinfo
   {journal} {Phys. Rev. D}\ }\textbf {\bibinfo {volume} {110}},\ \bibinfo
  {pages} {044002} (\bibinfo {year} {2024})},\ \Eprint
  {https://arxiv.org/abs/2404.16265} {arXiv:2404.16265 [gr-qc]} \BibitemShut
  {NoStop}%
\bibitem [{\citenamefont {Collaviti}\ \emph {et~al.}(2025)\citenamefont
  {Collaviti}, \citenamefont {Sun}, \citenamefont {Galanis},\ and\
  \citenamefont {Baryakhtar}}]{Collaviti:2024mvh}%
  \BibitemOpen
  \bibfield  {author} {\bibinfo {author} {\bibfnamefont {S.}~\bibnamefont
  {Collaviti}}, \bibinfo {author} {\bibfnamefont {L.}~\bibnamefont {Sun}},
  \bibinfo {author} {\bibfnamefont {M.}~\bibnamefont {Galanis}},\ and\ \bibinfo
  {author} {\bibfnamefont {M.}~\bibnamefont {Baryakhtar}},\ }\bibfield  {title}
  {\bibinfo {title} {{Observational prospects of self-interacting scalar
  superradiance with next-generation gravitational-wave detectors}},\ }\href
  {https://doi.org/10.1088/1361-6382/ad96ff} {\bibfield  {journal} {\bibinfo
  {journal} {Class. Quant. Grav.}\ }\textbf {\bibinfo {volume} {42}},\ \bibinfo
  {pages} {025006} (\bibinfo {year} {2025})},\ \Eprint
  {https://arxiv.org/abs/2407.04304} {arXiv:2407.04304 [gr-qc]} \BibitemShut
  {NoStop}%
\bibitem [{\citenamefont {Chen}\ \emph {et~al.}(2022)\citenamefont {Chen},
  \citenamefont {Liu}, \citenamefont {Lu}, \citenamefont {Mizuno},
  \citenamefont {Shu}, \citenamefont {Xue}, \citenamefont {Yuan},\ and\
  \citenamefont {Zhao}}]{Chen_2022}%
  \BibitemOpen
  \bibfield  {author} {\bibinfo {author} {\bibfnamefont {Y.}~\bibnamefont
  {Chen}}, \bibinfo {author} {\bibfnamefont {Y.}~\bibnamefont {Liu}}, \bibinfo
  {author} {\bibfnamefont {R.-S.}\ \bibnamefont {Lu}}, \bibinfo {author}
  {\bibfnamefont {Y.}~\bibnamefont {Mizuno}}, \bibinfo {author} {\bibfnamefont
  {J.}~\bibnamefont {Shu}}, \bibinfo {author} {\bibfnamefont {X.}~\bibnamefont
  {Xue}}, \bibinfo {author} {\bibfnamefont {Q.}~\bibnamefont {Yuan}},\ and\
  \bibinfo {author} {\bibfnamefont {Y.}~\bibnamefont {Zhao}},\ }\bibfield
  {title} {\bibinfo {title} {Stringent axion constraints with event horizon
  telescope polarimetric measurements of m87},\ }\href
  {https://doi.org/10.1038/s41550-022-01620-3} {\bibfield  {journal} {\bibinfo
  {journal} {Nature Astronomy}\ }\textbf {\bibinfo {volume} {6}},\ \bibinfo
  {pages} {592–598} (\bibinfo {year} {2022})}\BibitemShut {NoStop}%
\bibitem [{\citenamefont {Boskovic}\ \emph {et~al.}(2019)\citenamefont
  {Boskovic}, \citenamefont {Brito}, \citenamefont {Cardoso}, \citenamefont
  {Ikeda},\ and\ \citenamefont {Witek}}]{Boskovic:2018lkj}%
  \BibitemOpen
  \bibfield  {author} {\bibinfo {author} {\bibfnamefont {M.}~\bibnamefont
  {Boskovic}}, \bibinfo {author} {\bibfnamefont {R.}~\bibnamefont {Brito}},
  \bibinfo {author} {\bibfnamefont {V.}~\bibnamefont {Cardoso}}, \bibinfo
  {author} {\bibfnamefont {T.}~\bibnamefont {Ikeda}},\ and\ \bibinfo {author}
  {\bibfnamefont {H.}~\bibnamefont {Witek}},\ }\bibfield  {title} {\bibinfo
  {title} {{Axionic instabilities and new black hole solutions}},\ }\href
  {https://doi.org/10.1103/PhysRevD.99.035006} {\bibfield  {journal} {\bibinfo
  {journal} {Phys. Rev. D}\ }\textbf {\bibinfo {volume} {99}},\ \bibinfo
  {pages} {035006} (\bibinfo {year} {2019})},\ \Eprint
  {https://arxiv.org/abs/1811.04945} {arXiv:1811.04945 [gr-qc]} \BibitemShut
  {NoStop}%
\bibitem [{\citenamefont {Ikeda}\ \emph {et~al.}(2019)\citenamefont {Ikeda},
  \citenamefont {Brito},\ and\ \citenamefont {Cardoso}}]{Ikeda_2019}%
  \BibitemOpen
  \bibfield  {author} {\bibinfo {author} {\bibfnamefont {T.}~\bibnamefont
  {Ikeda}}, \bibinfo {author} {\bibfnamefont {R.}~\bibnamefont {Brito}},\ and\
  \bibinfo {author} {\bibfnamefont {V.}~\bibnamefont {Cardoso}},\ }\bibfield
  {title} {\bibinfo {title} {Blasts of light from axions},\ }\bibfield
  {journal} {\bibinfo  {journal} {Physical Review Letters}\ }\textbf {\bibinfo
  {volume} {122}},\ \href {https://doi.org/10.1103/physrevlett.122.081101}
  {10.1103/physrevlett.122.081101} (\bibinfo {year} {2019})\BibitemShut
  {NoStop}%
\bibitem [{\citenamefont {Harding}(2013)}]{harding2013neutronstarzoo}%
  \BibitemOpen
  \bibfield  {author} {\bibinfo {author} {\bibfnamefont {A.~K.}\ \bibnamefont
  {Harding}},\ }\href {https://arxiv.org/abs/1302.0869} {\bibinfo {title} {The
  neutron star zoo}} (\bibinfo {year} {2013}),\ \Eprint
  {https://arxiv.org/abs/1302.0869} {arXiv:1302.0869 [astro-ph.HE]}
  \BibitemShut {NoStop}%
\bibitem [{\citenamefont {Zel'Dovich}(1971)}]{zel1971generation}%
  \BibitemOpen
  \bibfield  {author} {\bibinfo {author} {\bibfnamefont {Y.~B.}\ \bibnamefont
  {Zel'Dovich}},\ }\bibfield  {title} {\bibinfo {title} {Generation of waves by
  a rotating body},\ }\href@noop {} {\bibfield  {journal} {\bibinfo  {journal}
  {Soviet Journal of Experimental and Theoretical Physics Letters}\ }\textbf
  {\bibinfo {volume} {14}},\ \bibinfo {pages} {180} (\bibinfo {year}
  {1971})}\BibitemShut {NoStop}%
\bibitem [{\citenamefont {Zel'Dovich}(1972)}]{zel1972amplification}%
  \BibitemOpen
  \bibfield  {author} {\bibinfo {author} {\bibfnamefont {Y.}~\bibnamefont
  {Zel'Dovich}},\ }\bibfield  {title} {\bibinfo {title} {Amplification of
  cylindrical electromagnetic waves reflected from a rotating body},\
  }\href@noop {} {\bibfield  {journal} {\bibinfo  {journal} {Soviet
  Physics-JETP}\ }\textbf {\bibinfo {volume} {35}},\ \bibinfo {pages} {1085}
  (\bibinfo {year} {1972})}\BibitemShut {NoStop}%
\bibitem [{\citenamefont {Brito}\ \emph {et~al.}(2020)\citenamefont {Brito},
  \citenamefont {Cardoso},\ and\ \citenamefont {Pani}}]{Brito_2020}%
  \BibitemOpen
  \bibfield  {author} {\bibinfo {author} {\bibfnamefont {R.}~\bibnamefont
  {Brito}}, \bibinfo {author} {\bibfnamefont {V.}~\bibnamefont {Cardoso}},\
  and\ \bibinfo {author} {\bibfnamefont {P.}~\bibnamefont {Pani}},\ }\href
  {https://doi.org/10.1007/978-3-030-46622-0} {\emph {\bibinfo {title}
  {Superradiance: New Frontiers in Black Hole Physics}}}\ (\bibinfo
  {publisher} {Springer International Publishing},\ \bibinfo {year}
  {2020})\BibitemShut {NoStop}%
\bibitem [{\citenamefont {Cardoso}\ \emph {et~al.}(2015)\citenamefont
  {Cardoso}, \citenamefont {Brito},\ and\ \citenamefont {Rosa}}]{Cardoso_2015}%
  \BibitemOpen
  \bibfield  {author} {\bibinfo {author} {\bibfnamefont {V.}~\bibnamefont
  {Cardoso}}, \bibinfo {author} {\bibfnamefont {R.}~\bibnamefont {Brito}},\
  and\ \bibinfo {author} {\bibfnamefont {J.~L.}\ \bibnamefont {Rosa}},\
  }\bibfield  {title} {\bibinfo {title} {Superradiance in stars},\ }\bibfield
  {journal} {\bibinfo  {journal} {Physical Review D}\ }\textbf {\bibinfo
  {volume} {91}},\ \href {https://doi.org/10.1103/physrevd.91.124026}
  {10.1103/physrevd.91.124026} (\bibinfo {year} {2015})\BibitemShut {NoStop}%
\bibitem [{\citenamefont {Cardoso}\ \emph {et~al.}(2017)\citenamefont
  {Cardoso}, \citenamefont {Pani},\ and\ \citenamefont {Yu}}]{Cardoso_2017}%
  \BibitemOpen
  \bibfield  {author} {\bibinfo {author} {\bibfnamefont {V.}~\bibnamefont
  {Cardoso}}, \bibinfo {author} {\bibfnamefont {P.}~\bibnamefont {Pani}},\ and\
  \bibinfo {author} {\bibfnamefont {T.-T.}\ \bibnamefont {Yu}},\ }\bibfield
  {title} {\bibinfo {title} {Superradiance in rotating stars and pulsar-timing
  constraints on dark photons},\ }\bibfield  {journal} {\bibinfo  {journal}
  {Physical Review D}\ }\textbf {\bibinfo {volume} {95}},\ \href
  {https://doi.org/10.1103/physrevd.95.124056} {10.1103/physrevd.95.124056}
  (\bibinfo {year} {2017})\BibitemShut {NoStop}%
\bibitem [{\citenamefont {Day}\ and\ \citenamefont
  {McDonald}(2019)}]{Day_2019}%
  \BibitemOpen
  \bibfield  {author} {\bibinfo {author} {\bibfnamefont {F.~V.}\ \bibnamefont
  {Day}}\ and\ \bibinfo {author} {\bibfnamefont {J.~I.}\ \bibnamefont
  {McDonald}},\ }\bibfield  {title} {\bibinfo {title} {Axion superradiance in
  rotating neutron stars},\ }\href
  {https://doi.org/10.1088/1475-7516/2019/10/051} {\bibfield  {journal}
  {\bibinfo  {journal} {Journal of Cosmology and Astroparticle Physics}\
  }\textbf {\bibinfo {volume} {2019}}\bibinfo  {number} { (10)},\ \bibinfo
  {pages} {051–051}}\BibitemShut {NoStop}%
\bibitem [{\citenamefont {Kaplan}\ \emph {et~al.}(2019)\citenamefont {Kaplan},
  \citenamefont {Rajendran},\ and\ \citenamefont {Riggins}}]{Kaplan:2019ako}%
  \BibitemOpen
\bibfield  {number} {  }\bibfield  {author} {\bibinfo {author} {\bibfnamefont
  {D.~E.}\ \bibnamefont {Kaplan}}, \bibinfo {author} {\bibfnamefont
  {S.}~\bibnamefont {Rajendran}},\ and\ \bibinfo {author} {\bibfnamefont
  {P.}~\bibnamefont {Riggins}},\ }\bibfield  {title} {\bibinfo {title}
  {{Particle Probes with Superradiant Pulsars}},\ }\href@noop {} {\  (\bibinfo
  {year} {2019})},\ \Eprint {https://arxiv.org/abs/1908.10440}
  {arXiv:1908.10440 [hep-ph]} \BibitemShut {NoStop}%
\bibitem [{\citenamefont {Chang}\ \emph {et~al.}(2025)\citenamefont {Chang},
  \citenamefont {Gao}, \citenamefont {Jaramillo}, \citenamefont {Meng},\ and\
  \citenamefont {Zhou}}]{Chang:2024xjp}%
  \BibitemOpen
  \bibfield  {author} {\bibinfo {author} {\bibfnamefont {F.-M.}\ \bibnamefont
  {Chang}}, \bibinfo {author} {\bibfnamefont {H.-Y.}\ \bibnamefont {Gao}},
  \bibinfo {author} {\bibfnamefont {V.}~\bibnamefont {Jaramillo}}, \bibinfo
  {author} {\bibfnamefont {X.}~\bibnamefont {Meng}},\ and\ \bibinfo {author}
  {\bibfnamefont {S.-Y.}\ \bibnamefont {Zhou}},\ }\bibfield  {title} {\bibinfo
  {title} {{Boson star superradiance with spinning effects and in time
  domain}},\ }\href {https://doi.org/10.1103/PhysRevD.111.044053} {\bibfield
  {journal} {\bibinfo  {journal} {Phys. Rev. D}\ }\textbf {\bibinfo {volume}
  {111}},\ \bibinfo {pages} {044053} (\bibinfo {year} {2025})},\ \Eprint
  {https://arxiv.org/abs/2412.01894} {arXiv:2412.01894 [gr-qc]} \BibitemShut
  {NoStop}%
\bibitem [{\citenamefont {Siemonsen}\ and\ \citenamefont
  {East}(2021)}]{Siemonsen_2021}%
  \BibitemOpen
  \bibfield  {author} {\bibinfo {author} {\bibfnamefont {N.}~\bibnamefont
  {Siemonsen}}\ and\ \bibinfo {author} {\bibfnamefont {W.~E.}\ \bibnamefont
  {East}},\ }\bibfield  {title} {\bibinfo {title} {Stability of rotating scalar
  boson stars with nonlinear interactions},\ }\bibfield  {journal} {\bibinfo
  {journal} {Physical Review D}\ }\textbf {\bibinfo {volume} {103}},\ \href
  {https://doi.org/10.1103/physrevd.103.044022} {10.1103/physrevd.103.044022}
  (\bibinfo {year} {2021})\BibitemShut {NoStop}%
\bibitem [{\citenamefont {Gao}\ \emph {et~al.}(2023)\citenamefont {Gao},
  \citenamefont {Saffin}, \citenamefont {Wang}, \citenamefont {Xie},\ and\
  \citenamefont {Zhou}}]{gao2023bosonstarsuperradiance}%
  \BibitemOpen
  \bibfield  {author} {\bibinfo {author} {\bibfnamefont {H.-Y.}\ \bibnamefont
  {Gao}}, \bibinfo {author} {\bibfnamefont {P.~M.}\ \bibnamefont {Saffin}},
  \bibinfo {author} {\bibfnamefont {Y.-J.}\ \bibnamefont {Wang}}, \bibinfo
  {author} {\bibfnamefont {Q.-X.}\ \bibnamefont {Xie}},\ and\ \bibinfo {author}
  {\bibfnamefont {S.-Y.}\ \bibnamefont {Zhou}},\ }\href
  {https://arxiv.org/abs/2306.01868} {\bibinfo {title} {Boson star
  superradiance}} (\bibinfo {year} {2023}),\ \Eprint
  {https://arxiv.org/abs/2306.01868} {arXiv:2306.01868 [gr-qc]} \BibitemShut
  {NoStop}%
\bibitem [{\citenamefont {Chadha-Day}\ \emph {et~al.}(2022)\citenamefont
  {Chadha-Day}, \citenamefont {Garbrecht},\ and\ \citenamefont
  {McDonald}}]{Chadha-Day:2022inf}%
  \BibitemOpen
  \bibfield  {author} {\bibinfo {author} {\bibfnamefont {F.}~\bibnamefont
  {Chadha-Day}}, \bibinfo {author} {\bibfnamefont {B.}~\bibnamefont
  {Garbrecht}},\ and\ \bibinfo {author} {\bibfnamefont {J.}~\bibnamefont
  {McDonald}},\ }\bibfield  {title} {\bibinfo {title} {{Superradiance in stars:
  non-equilibrium approach to damping of fields in stellar media}},\ }\href
  {https://doi.org/10.1088/1475-7516/2022/12/008} {\bibfield  {journal}
  {\bibinfo  {journal} {JCAP}\ }\textbf {\bibinfo {volume} {12}},\ \bibinfo
  {pages} {008}},\ \Eprint {https://arxiv.org/abs/2207.07662} {arXiv:2207.07662
  [hep-ph]} \BibitemShut {NoStop}%
\bibitem [{\citenamefont {Spieksma}\ and\ \citenamefont
  {Cannizzaro}(2025)}]{Spieksma:2025sda}%
  \BibitemOpen
  \bibfield  {author} {\bibinfo {author} {\bibfnamefont {T.~F.~M.}\
  \bibnamefont {Spieksma}}\ and\ \bibinfo {author} {\bibfnamefont
  {E.}~\bibnamefont {Cannizzaro}},\ }\bibfield  {title} {\bibinfo {title}
  {{Axion dissipation in conductive media and neutron star superradiance}},\
  }\href@noop {} {\  (\bibinfo {year} {2025})},\ \Eprint
  {https://arxiv.org/abs/2503.19978} {arXiv:2503.19978 [hep-ph]} \BibitemShut
  {NoStop}%
\bibitem [{\citenamefont {Brambilla}\ \emph {et~al.}(2015)\citenamefont
  {Brambilla}, \citenamefont {Kalapotharakos}, \citenamefont {Harding},\ and\
  \citenamefont {Kazanas}}]{Brambilla:2015vta}%
  \BibitemOpen
  \bibfield  {author} {\bibinfo {author} {\bibfnamefont {G.}~\bibnamefont
  {Brambilla}}, \bibinfo {author} {\bibfnamefont {C.}~\bibnamefont
  {Kalapotharakos}}, \bibinfo {author} {\bibfnamefont {A.~K.}\ \bibnamefont
  {Harding}},\ and\ \bibinfo {author} {\bibfnamefont {D.}~\bibnamefont
  {Kazanas}},\ }\bibfield  {title} {\bibinfo {title} {{Testing dissipative
  magnetosphere model light curves and spectra with FERMI pulsars}},\ }\href
  {https://doi.org/10.1088/0004-637X/804/2/84} {\bibfield  {journal} {\bibinfo
  {journal} {Astrophys. J.}\ }\textbf {\bibinfo {volume} {804}},\ \bibinfo
  {pages} {84} (\bibinfo {year} {2015})},\ \Eprint
  {https://arxiv.org/abs/1503.00744} {arXiv:1503.00744 [astro-ph.HE]}
  \BibitemShut {NoStop}%
\bibitem [{\citenamefont {Rezzolla}\ \emph {et~al.}(2001)\citenamefont
  {Rezzolla}, \citenamefont {Ahmedov},\ and\ \citenamefont
  {Miller}}]{Rezzolla_2001}%
  \BibitemOpen
  \bibfield  {author} {\bibinfo {author} {\bibfnamefont {L.}~\bibnamefont
  {Rezzolla}}, \bibinfo {author} {\bibfnamefont {B.~J.}\ \bibnamefont
  {Ahmedov}},\ and\ \bibinfo {author} {\bibfnamefont {J.~C.}\ \bibnamefont
  {Miller}},\ }\bibfield  {title} {\bibinfo {title} {General relativistic
  electromagnetic fields of a slowly rotating magnetized neutron star - i.
  formulation of the equations},\ }\href
  {https://doi.org/10.1046/j.1365-8711.2001.04161.x} {\bibfield  {journal}
  {\bibinfo  {journal} {Monthly Notices of the Royal Astronomical Society}\
  }\textbf {\bibinfo {volume} {322}},\ \bibinfo {pages} {723–740} (\bibinfo
  {year} {2001})}\BibitemShut {NoStop}%
\bibitem [{\citenamefont {Andersson}\ and\ \citenamefont
  {Kokkotas}(2001)}]{Andersson:2000mf}%
  \BibitemOpen
  \bibfield  {author} {\bibinfo {author} {\bibfnamefont {N.}~\bibnamefont
  {Andersson}}\ and\ \bibinfo {author} {\bibfnamefont {K.~D.}\ \bibnamefont
  {Kokkotas}},\ }\bibfield  {title} {\bibinfo {title} {{The R mode instability
  in rotating neutron stars}},\ }\href
  {https://doi.org/10.1142/S0218271801001062} {\bibfield  {journal} {\bibinfo
  {journal} {Int. J. Mod. Phys. D}\ }\textbf {\bibinfo {volume} {10}},\
  \bibinfo {pages} {381} (\bibinfo {year} {2001})},\ \Eprint
  {https://arxiv.org/abs/gr-qc/0010102} {arXiv:gr-qc/0010102} \BibitemShut
  {NoStop}%
\bibitem [{\citenamefont {Manchester}\ \emph {et~al.}(2005)\citenamefont
  {Manchester}, \citenamefont {Hobbs}, \citenamefont {Teoh},\ and\
  \citenamefont {Hobbs}}]{Manchester_2005}%
  \BibitemOpen
  \bibfield  {author} {\bibinfo {author} {\bibfnamefont {R.~N.}\ \bibnamefont
  {Manchester}}, \bibinfo {author} {\bibfnamefont {G.~B.}\ \bibnamefont
  {Hobbs}}, \bibinfo {author} {\bibfnamefont {A.}~\bibnamefont {Teoh}},\ and\
  \bibinfo {author} {\bibfnamefont {M.}~\bibnamefont {Hobbs}},\ }\bibfield
  {title} {\bibinfo {title} {The australia telescope national facility pulsar
  catalogue},\ }\href {https://doi.org/10.1086/428488} {\bibfield  {journal}
  {\bibinfo  {journal} {The Astronomical Journal}\ }\textbf {\bibinfo {volume}
  {129}},\ \bibinfo {pages} {1993–2006} (\bibinfo {year} {2005})}\BibitemShut
  {NoStop}%
\bibitem [{\citenamefont {Romani}\ \emph {et~al.}(2022)\citenamefont {Romani},
  \citenamefont {Kandel}, \citenamefont {Filippenko}, \citenamefont {Brink},\
  and\ \citenamefont {Zheng}}]{Romani_2022}%
  \BibitemOpen
  \bibfield  {author} {\bibinfo {author} {\bibfnamefont {R.~W.}\ \bibnamefont
  {Romani}}, \bibinfo {author} {\bibfnamefont {D.}~\bibnamefont {Kandel}},
  \bibinfo {author} {\bibfnamefont {A.~V.}\ \bibnamefont {Filippenko}},
  \bibinfo {author} {\bibfnamefont {T.~G.}\ \bibnamefont {Brink}},\ and\
  \bibinfo {author} {\bibfnamefont {W.}~\bibnamefont {Zheng}},\ }\bibfield
  {title} {\bibinfo {title} {Psr j0952-0607: The fastest and heaviest known
  galactic neutron star},\ }\href {https://doi.org/10.3847/2041-8213/ac8007}
  {\bibfield  {journal} {\bibinfo  {journal} {The Astrophysical Journal
  Letters}\ }\textbf {\bibinfo {volume} {934}},\ \bibinfo {pages} {L17}
  (\bibinfo {year} {2022})}\BibitemShut {NoStop}%
\bibitem [{\citenamefont {G\"artlein}\ \emph {et~al.}(2024)\citenamefont
  {G\"artlein}, \citenamefont {Sagun}, \citenamefont {Ivanytskyi},
  \citenamefont {Blaschke},\ and\ \citenamefont {Lopes}}]{Gartlein:2024cbj}%
  \BibitemOpen
  \bibfield  {author} {\bibinfo {author} {\bibfnamefont {C.}~\bibnamefont
  {G\"artlein}}, \bibinfo {author} {\bibfnamefont {V.}~\bibnamefont {Sagun}},
  \bibinfo {author} {\bibfnamefont {O.}~\bibnamefont {Ivanytskyi}}, \bibinfo
  {author} {\bibfnamefont {D.}~\bibnamefont {Blaschke}},\ and\ \bibinfo
  {author} {\bibfnamefont {I.}~\bibnamefont {Lopes}},\ }\bibfield  {title}
  {\bibinfo {title} {{Fastest spinning millisecond pulsars: indicators for
  quark matter in neutron stars?}},\ }\href@noop {} {\  (\bibinfo {year}
  {2024})},\ \Eprint {https://arxiv.org/abs/2412.07758} {arXiv:2412.07758
  [nucl-th]} \BibitemShut {NoStop}%
\bibitem [{\citenamefont {Kirk}\ \emph {et~al.}(2009)\citenamefont {Kirk},
  \citenamefont {Lyubarsky},\ and\ \citenamefont {Petri}}]{Kirk:2007tn}%
  \BibitemOpen
  \bibfield  {author} {\bibinfo {author} {\bibfnamefont {J.~G.}\ \bibnamefont
  {Kirk}}, \bibinfo {author} {\bibfnamefont {Y.}~\bibnamefont {Lyubarsky}},\
  and\ \bibinfo {author} {\bibfnamefont {J.}~\bibnamefont {Petri}},\ }\bibfield
   {title} {\bibinfo {title} {{The theory of pulsar winds and nebulae}},\
  }\href {https://doi.org/10.1007/978-3-540-76965-1_16} {\bibfield  {journal}
  {\bibinfo  {journal} {Astrophys. Space Sci. Libr.}\ }\textbf {\bibinfo
  {volume} {357}},\ \bibinfo {pages} {421} (\bibinfo {year} {2009})},\ \Eprint
  {https://arxiv.org/abs/astro-ph/0703116} {arXiv:astro-ph/0703116}
  \BibitemShut {NoStop}%
\bibitem [{\citenamefont {Cerutti}\ and\ \citenamefont
  {Giacinti}(2020)}]{Cerutti:2020qav}%
  \BibitemOpen
  \bibfield  {author} {\bibinfo {author} {\bibfnamefont {B.}~\bibnamefont
  {Cerutti}}\ and\ \bibinfo {author} {\bibfnamefont {G.}~\bibnamefont
  {Giacinti}},\ }\bibfield  {title} {\bibinfo {title} {{A global model of
  particle acceleration at pulsar wind termination shocks}},\ }\href
  {https://doi.org/10.1051/0004-6361/202038883} {\bibfield  {journal} {\bibinfo
   {journal} {Astron. Astrophys.}\ }\textbf {\bibinfo {volume} {642}},\
  \bibinfo {pages} {A123} (\bibinfo {year} {2020})},\ \Eprint
  {https://arxiv.org/abs/2008.07253} {arXiv:2008.07253 [astro-ph.HE]}
  \BibitemShut {NoStop}%
\bibitem [{\citenamefont {{Michel}}(1969)}]{1969ApJ...158..727M}%
  \BibitemOpen
  \bibfield  {author} {\bibinfo {author} {\bibfnamefont {F.~C.}\ \bibnamefont
  {{Michel}}},\ }\bibfield  {title} {\bibinfo {title} {{Relativistic
  Stellar-Wind Torques}},\ }\href {https://doi.org/10.1086/150233} {\bibfield
  {journal} {\bibinfo  {journal} {\apj}\ }\textbf {\bibinfo {volume} {158}},\
  \bibinfo {pages} {727} (\bibinfo {year} {1969})}\BibitemShut {NoStop}%
\bibitem [{\citenamefont {{P{\'e}tri}}\ \emph {et~al.}(2002)\citenamefont
  {{P{\'e}tri}}, \citenamefont {{Heyvaerts}},\ and\ \citenamefont
  {{Bonazzola}}}]{2002A&A...384..414P}%
  \BibitemOpen
  \bibfield  {author} {\bibinfo {author} {\bibfnamefont {J.}~\bibnamefont
  {{P{\'e}tri}}}, \bibinfo {author} {\bibfnamefont {J.}~\bibnamefont
  {{Heyvaerts}}},\ and\ \bibinfo {author} {\bibfnamefont {S.}~\bibnamefont
  {{Bonazzola}}},\ }\bibfield  {title} {\bibinfo {title} {{Global static
  electrospheres of charged pulsars}},\ }\href
  {https://doi.org/10.1051/0004-6361:20020044} {\bibfield  {journal} {\bibinfo
  {journal} {aap}\ }\textbf {\bibinfo {volume} {384}},\ \bibinfo {pages} {414}
  (\bibinfo {year} {2002})}\BibitemShut {NoStop}%
\bibitem [{\citenamefont {Gu\'epin}\ \emph {et~al.}(2020)\citenamefont
  {Gu\'epin}, \citenamefont {Cerutti},\ and\ \citenamefont
  {Kotera}}]{Guepin:2019fjb}%
  \BibitemOpen
  \bibfield  {author} {\bibinfo {author} {\bibfnamefont {C.}~\bibnamefont
  {Gu\'epin}}, \bibinfo {author} {\bibfnamefont {B.}~\bibnamefont {Cerutti}},\
  and\ \bibinfo {author} {\bibfnamefont {K.}~\bibnamefont {Kotera}},\
  }\bibfield  {title} {\bibinfo {title} {{Proton acceleration in pulsar
  magnetospheres}},\ }\href {https://doi.org/10.1051/0004-6361/201936816}
  {\bibfield  {journal} {\bibinfo  {journal} {Astron. Astrophys.}\ }\textbf
  {\bibinfo {volume} {635}},\ \bibinfo {pages} {A138} (\bibinfo {year}
  {2020})},\ \Eprint {https://arxiv.org/abs/1910.11387} {arXiv:1910.11387
  [astro-ph.HE]} \BibitemShut {NoStop}%
\bibitem [{\citenamefont {Cerutti}\ and\ \citenamefont
  {Beloborodov}(2017)}]{Cerutti:2016ttn}%
  \BibitemOpen
  \bibfield  {author} {\bibinfo {author} {\bibfnamefont {B.}~\bibnamefont
  {Cerutti}}\ and\ \bibinfo {author} {\bibfnamefont {A.}~\bibnamefont
  {Beloborodov}},\ }\bibfield  {title} {\bibinfo {title} {{Electrodynamics of
  pulsar magnetospheres}},\ }\href {https://doi.org/10.1007/s11214-016-0315-7}
  {\bibfield  {journal} {\bibinfo  {journal} {Space Sci. Rev.}\ }\textbf
  {\bibinfo {volume} {207}},\ \bibinfo {pages} {111} (\bibinfo {year}
  {2017})},\ \Eprint {https://arxiv.org/abs/1611.04331} {arXiv:1611.04331
  [astro-ph.HE]} \BibitemShut {NoStop}%
\bibitem [{\citenamefont {Haskell}\ \emph {et~al.}(2018)\citenamefont
  {Haskell}, \citenamefont {Zdunik}, \citenamefont {Fortin}, \citenamefont
  {Bejger}, \citenamefont {Wijnands},\ and\ \citenamefont
  {Patruno}}]{Haskell:2018nlh}%
  \BibitemOpen
  \bibfield  {author} {\bibinfo {author} {\bibfnamefont {B.}~\bibnamefont
  {Haskell}}, \bibinfo {author} {\bibfnamefont {J.~L.}\ \bibnamefont {Zdunik}},
  \bibinfo {author} {\bibfnamefont {M.}~\bibnamefont {Fortin}}, \bibinfo
  {author} {\bibfnamefont {M.}~\bibnamefont {Bejger}}, \bibinfo {author}
  {\bibfnamefont {R.}~\bibnamefont {Wijnands}},\ and\ \bibinfo {author}
  {\bibfnamefont {A.}~\bibnamefont {Patruno}},\ }\bibfield  {title} {\bibinfo
  {title} {{Fundamental physics and the absence of sub-millisecond pulsars}},\
  }\href {https://doi.org/10.1051/0004-6361/201833521} {\bibfield  {journal}
  {\bibinfo  {journal} {Astron. Astrophys.}\ }\textbf {\bibinfo {volume}
  {620}},\ \bibinfo {pages} {A69} (\bibinfo {year} {2018})},\ \Eprint
  {https://arxiv.org/abs/1805.11277} {arXiv:1805.11277 [astro-ph.HE]}
  \BibitemShut {NoStop}%
\bibitem [{\citenamefont {\c{C}\i{}k\i{}nto\u{g}lu}\ and\ \citenamefont
  {Ek\c{s}i}(2023)}]{Cikintoglu:2023drm}%
  \BibitemOpen
  \bibfield  {author} {\bibinfo {author} {\bibfnamefont {S.}~\bibnamefont
  {\c{C}\i{}k\i{}nto\u{g}lu}}\ and\ \bibinfo {author} {\bibfnamefont {K.~Y.}\
  \bibnamefont {Ek\c{s}i}},\ }\bibfield  {title} {\bibinfo {title} {{On the
  minimum spin period of accreting pulsars}},\ }\href
  {https://doi.org/10.1093/mnras/stad2036} {\bibfield  {journal} {\bibinfo
  {journal} {Mon. Not. Roy. Astron. Soc.}\ }\textbf {\bibinfo {volume} {524}},\
  \bibinfo {pages} {4899} (\bibinfo {year} {2023})},\ \Eprint
  {https://arxiv.org/abs/2302.10649} {arXiv:2302.10649 [astro-ph.HE]}
  \BibitemShut {NoStop}%
\bibitem [{\citenamefont {Liu}\ \emph {et~al.}(2023)\citenamefont {Liu},
  \citenamefont {Lou},\ and\ \citenamefont {Ren}}]{Liu:2021zlt}%
  \BibitemOpen
  \bibfield  {author} {\bibinfo {author} {\bibfnamefont {T.}~\bibnamefont
  {Liu}}, \bibinfo {author} {\bibfnamefont {X.}~\bibnamefont {Lou}},\ and\
  \bibinfo {author} {\bibfnamefont {J.}~\bibnamefont {Ren}},\ }\bibfield
  {title} {\bibinfo {title} {{Pulsar Polarization Arrays}},\ }\href
  {https://doi.org/10.1103/PhysRevLett.130.121401} {\bibfield  {journal}
  {\bibinfo  {journal} {Phys. Rev. Lett.}\ }\textbf {\bibinfo {volume} {130}},\
  \bibinfo {pages} {121401} (\bibinfo {year} {2023})},\ \Eprint
  {https://arxiv.org/abs/2111.10615} {arXiv:2111.10615 [astro-ph.HE]}
  \BibitemShut {NoStop}%
\end{thebibliography}%

\end{document}